%% file: alicepreprint_CDS_20170731.tex
\documentclass[ALICE,manyauthors]{cernphprep}

\usepackage[comma,square,numbers,sort&compress]{natbib}
\usepackage{hyperref}
\usepackage{lineno}
\usepackage{graphicx,subfigure}
\usepackage[utf8]{inputenc}
   
\usepackage{color}
\definecolor{dgreen}{cmyk}{1.,0.,1.,0.2}        
\definecolor{orange}{cmyk}{0.,0.353,1.,0.}    

\def\snn{\mbox{$\sqrt{s_{_{\rm NN}}}$}}
\def\pt{p_{\rm{T}}}

\begin{document}%

\begin{titlepage}
\PHyear{2017}
\PHnumber{149}      
\PHdate{29 Jun}  
%

\title{Searches for transverse momentum dependent flow vector fluctuations\\ in Pb--Pb and p--Pb collisions at the LHC}
\ShortTitle{Flow vector fluctuations in Pb--Pb and p--Pb collisions}   

\Collaboration{ALICE Collaboration\thanks{See Appendix~\ref{app:collab} for the list of collaboration members}}
\ShortAuthor{ALICE Collaboration} 

\begin{abstract} 
The measurement of azimuthal correlations of charged particles is presented for Pb--Pb collisions at $\sqrt{s_{_{\rm NN}}}=$ 2.76 TeV and p--Pb collisions at $\sqrt{s_{_{\rm NN}}}=$ 5.02 TeV with the ALICE detector at the CERN Large Hadron Collider. 
These correlations are measured for the second, third and fourth order flow vector in the pseudorapidity region $|\eta|<0.8$ as a function of centrality and transverse momentum $p_{\rm T}$ using two observables, to search for evidence of $p_{\rm T}$-dependent flow vector fluctuations.
For Pb--Pb collisions at 2.76 TeV, the measurements indicate that $p_{\rm T}$-dependent fluctuations are only present for the second order flow vector. Similar results have been found for p--Pb collisions at 5.02 TeV. These measurements are compared to hydrodynamic model calculations with event-by-event geometry fluctuations in the initial state to constrain the initial conditions and transport properties of the matter created in Pb--Pb and p--Pb collisions.  
\end{abstract}
\end{titlepage}
\setcounter{page}{2}


\section{Introduction}
\label{sec:intro}
 
The primary goal of ultrarelativistic heavy-ion collisions is to study the properties of the Quark-Gluon Plasma (QGP), a state of matter predicted by Quantum Chromodynamics to exist at high temperatures and energy densities~\cite{Shuryak:1978ij, Shuryak:1980tp}. An important experimental observable used to accomplish this goal is the azimuthal anisotropy of particles emitted in the transverse plane. 
In non-central heavy-ion collisions, the overlap region of the Lorentz-contracted nuclei is roughly almond-shaped.
Nucleons contained in such anisotropic overlap region interact with each other and give rise to a system of high energy density which expands anisotropically. 
These interactions convert the initial spatial asymmetry into a final-state momentum anisotropy of the produced particles, a phenomenon referred to as collective anisotropic flow~\cite{Ollitrault:1992bk,Voloshin:2008dg, Heinz:2013th}. 
Anisotropic flow is characterised using a Fourier decomposition of the azimuthal distribution of particles with respect to the flow symmetry planes~\cite{Voloshin:1994mz, Poskanzer:1998yz} 
\begin{equation}
E\frac{{\rm d}^{3}N}{{\rm d}^{3}{\vec{p}}} = \frac{1}{2\pi} \frac{{\rm d}^{2}N}{p_{\rm T}{\rm d}p_{\rm T}{\rm d}\eta}(1+2\sum_{n=1}^{\infty}v_{n} \cos[n(\varphi - \Psi_{n})]),
\label{eq:FourierExp}
\end{equation}
where $N$ is the number of produced particles, $E$ is the energy, $\vec{p}$ the momentum, $p_{\rm T}$ the transverse momentum, $\varphi$ the azimuthal angle and $\eta$ the pseudorapidity of the particle. 
The $n^{th}$ order flow (vector) $V_{n}$ is defined as: $V_{n} \equiv v_{n} \, e^{in\Psi_{n}}$, where $v_{n}$ is the flow coefficient, and $\Psi_{n}$ represents the azimuth of $V_{n}$ in momentum space (flow angle).
For a uniform matter distribution in the initial stage of a heavy-ion collision, $\Psi_{n}$ for $n \geq 1$ coincides with the reaction plane defined by the beam direction and impact parameter. Due to event-by-event fluctuations of the participating nucleons distribution inside the overlap region, the $\Psi_{n}$ may deviate from the reaction plane and the odd flow coefficients $v_{2n-1}$ are non-vanishing~\cite{Mishra:2007tw, Mishra:2008dm, Takahashi:2009na, Alver:2010gr, Alver:2010dn, Teaney:2010vd, Luzum:2010sp}. Large flow coefficients were observed at the Relativistic Heavy-Ion Collider (RHIC)~\cite{Arsene:2004fa, Back:2004je, Adams:2005dq, Adcox:2004mh} and the Large Hadron Collider (LHC)~\cite{Aamodt:2010pa,ALICE:2011ab, Abelev:2014pua, Adam:2016izf, Acharya:2017zfg, ATLAS:2011ah, ATLAS:2012at, Aad:2013xma, Chatrchyan:2012wg, Chatrchyan:2012ta, Chatrchyan:2012xq}. These measurements constrain the initial conditions (e.g. energy and entropy density) and transport coefficients of the system (such as shear viscosity over entropy density ratio, $\eta/s$). The recent measurements of correlations between different order flow coefficients and flow angles~\cite{ALICE:2016kpq,Acharya:2017zfg}, together with the comparisons to theoretical calculations, indicate that the matter created in ultrarelativistic heavy-ion collisions behaves as a nearly perfect fluid (almost zero $\eta/s$) whose constituent particles interact strongly~\cite{Song:2017wtw}.

Traditionally the final-state symmetry plane angles are estimated event-by-event from the particle azimuthal distribution over a large range in $\pt$. However, hydrodynamic calculations indicate a $\pt$ dependence of the flow vector $V_{n}$ due to event-by-event fluctuations in the initial energy density of the nuclear collisions~\cite{Gardim:2012im, Heinz:2013bua}. These flow vector fluctuations could be responsible for the experimentally observed breakdown of the factorisation~\cite{ATLAS:2012at, Chatrchyan:2012wg, Aamodt:2011by}. They might also affect the measured $\pt$-differential anisotropic flow $v_{n}(\pt)$~\cite{Heinz:2013bua}. 
Therefore, searches for $p_{\rm T}$-dependent flow vector fluctuations become important and these measurements together with the comparisons to theoretical calculations not only constrain the transport properties, but also shed light on the initial conditions in heavy-ion collisions. 

Studies of azimuthal correlations are performed also in p--Pb collisions at the LHC. The original goal of p--Pb collisions was to provide reference data for the high energy Pb--Pb collisions. However, indications of collective behaviour have been discovered by the ALICE, ATLAS, CMS and LHCb Collaborations~\cite{Abelev:2012ola,ABELEV:2013wsa,Abelev:2013haa,Abelev:2014mda,Aad:2012gla,Aad:2013fja,CMS:2013bza,Chatrchyan:2013nka,Chatrchyan:2013eya,Khachatryan:2014jra,Khachatryan:2015waa,Aaij:2015qcq}. If the azimuthal correlations in small collision systems reveal the onset of hydrodynamic flow behaviour, the breakdown of factorisation should be expected in small collision systems and reproduced by hydrodynamic calculations as well. 

The first experimental indication of $p_{\rm T}$-dependent flow vector fluctuations was observed by ALICE in studies of the decomposition of Fourier harmonics of the two-particle azimuthal correlations~\cite{Aamodt:2011by}. Fits to the azimuthal correlations, assuming factorisation of the two-particle Fourier harmonics, agree well with data up to $\pt^{\, a} \sim$ 3-4 GeV/$c$, deviations at higher $\pt$ are interpreted, as at least partially, due to away-side recoil jet contributions~\cite{Aamodt:2011by}. A systematic study of the factorisation of long-range two-particle Fourier harmonic into the flow coefficients is also performed in both Pb--Pb and p--Pb collisions by CMS~\cite{CMS:2013bza,Khachatryan:2015oea}. 

In this paper, the $p_{\rm T}$-dependent flow vector fluctuations are investigated in more detail using novel observables for azimuthal correlations, for charged particles in Pb--Pb collisions at $\snn=$ 2.76 TeV and p--Pb collisions at $\snn=$ 5.02 TeV with the ALICE detector. The definitions of the observables are given in Section~\ref{sec:methods}. 
The experimental setup is described in Section~\ref{sec:setup}.
The results are reported in multiple centrality classes for Pb--Pb collisions and multiplicity classes for p--Pb collisions for several transverse momentum intervals. Details of the event and track selections are given in Section~\ref{sec:evTrack}. Section~\ref{sec:sys} shows the study of systematic uncertainties of the aforementioned observables. Section~\ref{sec:Results} presents results and discussions while section~\ref{sec:summary} summarizes and concludes this work.



\section{Probes of $\mathbf{\emph p_\mathrm{\mathbf{T}}}$-dependent flow vector fluctuations}
\label{sec:methods}

The traditional approach used to measure anisotropic azimuthal correlations is as follows: first, the flow coefficient of reference particles (RPs), called reference flow, is determined over a wide kinematic range, and then the transverse momentum differential flow coefficient is calculated by correlating the particles of interest (POIs) with respect to the reference flow obtained in the first step. Usually a pseudorapidity gap $|\Delta\eta|$ is applied between the two correlated particles to suppress non-flow effects, which comprise azimuthal correlations not associated with flow symmetry planes, e.g. resonance decays and jet contributions. This approach has commonly been used to measure the anisotropic flow at the LHC~\cite{ALICE:2011ab,ATLAS:2012at,Chatrchyan:2012ta}. Considering possible $p_{\rm T}$-dependent flow angle and/or magnitude fluctuations and neglecting non-flow contributions, the flow coefficient from $\pt$ interval $a$ measured with 2-particle correlations can be expressed as
\begin{eqnarray}
\begin{split}
v_{n}\{2\}(\pt^{a})   = \frac{\langle \langle \cos \,[ \, n \, ( \varphi_{1}^{\, a} - \varphi_{2}^{\, \rm ref})] \rangle \rangle} {\sqrt{\langle \langle \cos \,[ \, n \, ( \varphi_{1}^{\, \rm ref} - \varphi_{2}^{\, \rm ref})] \rangle \rangle} } 
= \frac{ \langle v_{n}(\pt^{\, a}) \, v_{n}^{\, \rm ref} \, \cos \,[ \, n \, (\Psi_{n}(\pt^{\, a}) - \Psi_{n}) ]  \rangle} {\sqrt{ \langle {v_{n}^{\, \rm ref}}^{\, 2}  \rangle}}. 
\end{split}
\label{eq:v22modi}
\end{eqnarray}
Here, a single set of angular brackets denotes averaging over events, and a double set indicates averaging over both particles and events. The $\varphi^{\rm ref}$ and $\varphi^{a}$ represent the azimuthal angle of RPs and POIs, respectively. The $v_{n}^{\, \rm ref}$ stands for the reference flow, and $\Psi_n(\pt^{a})$ denotes the $\pt$ differential symmetry plane angle at $\pt^{a}$, which might fluctuate around the $\pt$ integrated symmetry plane angle $\Psi_{n}$. The cosine term $\langle \cos \,[ \, n \, (\Psi_{n}(\pt^{\, a}) - \Psi_{n}) ]  \rangle$ shows the effects of the difference between $\Psi_{n}(\pt)$ and $\Psi_{n}$, due to the $p_{\rm T}$-dependent flow angle fluctuations. Additionally, $\langle v_{n}(\pt^{a}) \, v_{n}^{\rm ref} \rangle$ cannot be factorised into the product of $\sqrt{\langle v_{n}(\pt^{\, a})^{2} \rangle}$ and $\sqrt{\langle {v_{n}^{\, \rm ref}}^{\, 2} \rangle}$ if there are $p_{\rm T}$-dependent flow coefficient fluctuations.

A new type of two-particle azimuthal correlations from $p_{\rm T}^{\, a}$, denoted as $v_{n}[2](\pt^a)$, is proposed in~\cite{Heinz:2013bua}:
\begin{equation}
\label{eq:newv22}
\begin{split}
v_{n}[2](\pt^{\, a}) = & \sqrt{ \langle \langle \cos \,[ \, n \, ( \varphi_{1}^{\, a} - \varphi_{2}^{\, a})]\rangle \rangle } \\
 = & \sqrt{ \langle \langle \cos \,[ \, n  ( \varphi_{1}^{\, a} - \Psi_{n}(\pt^{\, a})) - n \, ( \varphi_{2}^{\, a} - \Psi_{n}(\pt^{\, a})) ] \rangle \rangle }  \\
 =  & \sqrt{ \langle v_{n}(\pt^{\, a})^{2} \rangle } .
\end{split}
\end{equation}
The difference between $v_{n}\{2\}(\pt^a)$ and $v_{n}[2](\pt^a)$ is that the former takes the flow of RPs from a wide $p_{\rm T}$ range and the POIs from a certain $p_{\rm T}$ interval, while the latter is essentially the reference flow calculated within a narrow $\pt$ range.
The ratio of $v_{n}\{2\}$ and $v_{n}[2]$ allows $p_{\rm T}$-dependent flow vector fluctuations
\begin{equation}
\label{eq:v22EQ}
\frac{v_{n}\{2\}}{v_{n}[2]} (\pt^{\, a})  =  \frac{\langle v_{n}(\pt^{\, a}) \, v_{n}^{\, \rm ref} \, \cos \,[ \, n \, ( \Psi_{n}(\pt^{\, a}) - \Psi_{n})] \rangle }{ \sqrt{\langle v_{n}(\pt^{\, a})^{\, 2} \rangle }  \,  \sqrt{\langle {v_{n}^{\, \rm ref}}^{\, 2}  \rangle } } .
\end{equation}
When the correlations are dominated by flow, a ratio value smaller than unity shall indicate the presence of $p_{\rm T}$-dependent flow vector fluctuations.

Another observable to probe the $p_{\rm T}$-dependent flow vector fluctuations is the factorisation ratio $r_{n}$~\cite{Gardim:2012im, Heinz:2013bua}. 
It can be calculated using the two-particle Fourier harmonic as
\begin{equation}
r_{n} = \frac{V_{n\Delta}(\pt^{\, a}, \pt^{\, t})}{\sqrt{V_{n\Delta}(\pt^{\, a}, \pt^{\, a}) \, V_{n\Delta}(\pt^{\, t}, \pt^{\, t})}}, 
\end{equation}
where $V_{n\Delta}(\pt^{\, a}, \pt^{\, t})$ is the $n^{th}$ -order Fourier harmonic of the two-particle azimuthal correlations of triggered and associated particles from $\pt^{\, t}$ and $\pt^{\, a}$, and is calculated as 
\begin{equation}
V_{n\Delta}(\pt^{\, a}, \pt^{\, t}) =   \langle \langle \cos \,[ \, n \, ( \varphi_{1}^{\, a} - \varphi_{2}^{\, t})]\rangle \rangle = \langle v_n(\pt^{\, a}) \, v_n(\pt^{\, t}) \, \cos \,[ \, n(\Psi_n(\pt^{\, a}) - \Psi_n(\pt^{\, t}))]  \rangle,
\label{eq:Vndelta}
\end{equation}
where $\Psi_n(\pt^{\, a})$ and $\Psi_n(\pt^{\, t})$ represent the flow angles at $\pt^{\, a}$ and $\pt^{\, t}$, respectively. The subscript $_{\Delta}$ indicates that a pseudorapidity gap is usually applied to minimise contamination from non-flow effects. If both triggered and associated particle are from the same $\pt$ interval $\pt^t$, Eq.(\ref{eq:Vndelta}) reduces to
\begin{equation}
V_{n\Delta}(\pt^{\, a}, \pt^{\, a}) =   \langle \langle \cos \,[ \, n \, ( \varphi_{1}^{\, a} - \varphi_{2}^{\, a})]\rangle \rangle =  \langle v_n(\pt^{\, a})^{\, 2} \rangle.
\end{equation}
Similarly, we have
\begin{equation}
V_{n\Delta}(\pt^{\, t}, \pt^{\, t}) =   \langle \langle \cos \,[ \, n \, ( \varphi_{1}^{\, t} - \varphi_{2}^{\, t})]\rangle \rangle =  \langle v_n(\pt^{\, t})^{\, 2} \rangle
\end{equation}

In the end $r_n$ is equivalent to
\begin{equation}
r_n = \frac{\langle v_n(\pt^{\, a}) \, v_n(\pt^{\, t}) \, \cos \,[ \, n(\Psi_n(\pt^{\, a}) - \Psi_n(\pt^{\, t}))]  \rangle}{\sqrt{\langle v_n(\pt^{\, a})^{\, 2} \rangle  \langle v_n(\pt^{\, t})^{\, 2} \rangle}}.
\label{eq:rn_fact}
\end{equation}  
It can be seen that $r_{n} = 1$ does not always hold true, i.e. most of the known sources of non-flow effects do not factorise at low $\pt$, which is confirmed by Monte Carlo studies~\cite{Xu:2012ue}.
In a flow-dominated system, $r_n \le 1$ due to the Cauchy--Schwarz inequality. Factorisation implies $r_n=1$, while $r_n<1$ shows the breaking of factorisation, suggesting the presence of $p_{\rm T}$-dependent flow vector fluctuations~\cite{Gardim:2012im, Heinz:2013bua}.

Note that Eqs.~(\ref{eq:v22EQ}) and~(\ref{eq:rn_fact}) look very similar. The ratios $v_{n}\{2\} / v_{n}[2]$ include the $\pt$ integrated information and probe the $\pt$-differential flow vector with respect to the $\pt$ integrated flow vector. The $r_{n}$ carries more detailed information on the 2-particle correlation structure for triggered and associated particle from narrow $\pt$ intervals, and probe the fluctuations of flow vector at $\pt^a$ and $\pt^t$; however, it also has larger statistical uncertainties. If the triggered particles are selected from a very wide kinematic range, the observable $r_n$ becomes identical with $v_{n}\{2\} / v_{n}[2]$. In this paper, we study $v_{n}\{2\} / v_{n}[2]$ up to $n=4$ and $r_{n}$ up to $n=3$.

\section{Experimental setup}
\label{sec:setup}

A Large Ion Collider Experiment (ALICE)~\cite{Aamodt:2008zz} is the dedicated heavy-ion experiment at the LHC designed to study strongly interacting matter at extreme energy densities. It was built to cope with the large charged-particle multiplicity density in central Pb--Pb collisions at the LHC, with several thousand tracks per unit of pseudorapidity. The ALICE apparatus consists of a central barrel that measures hadrons, electrons, muons and photons, and a forward spectrometer for the identification of muons. Several smaller detectors in the forward region are used for triggering and global event characterization. The central barrel is located inside a solenoidal magnet that provides a magnetic field of up to 0.5 T. 
Charged tracks are reconstructed using the Time Projection Chamber (TPC)~\cite{Aamodt:2008zz, Alme:2010ke} and the Inner Tracking System (ITS)~\cite{Aamodt:2008zz, Aamodt:2010aa} with a track momentum resolution better than 2\% for the momentum range 0.2 $\textless~\pt~\textless$ 5.0 GeV/$c$~\cite{Abelev:2014ffa}. The TPC is the main tracking detector of the central barrel, sufficient with full azimuthal coverage in the range of $|\eta|<0.9$. The ITS consists of six layers of silicon detectors placed at radii between 3.9 cm and 43 cm and matching the pseudorapidity acceptance of the TPC. Three different technologies are employed in the ITS: the two innermost layers are equipped with Silicon Pixel Detectors (SPD), the following two layers have Silicon Drift Detectors (SDD) and the two outer layers are double-sided Silicon Strip Detectors (SSD).
The V0 detector~\cite{Aamodt:2008zz, Abbas:2013taa} was used for triggering and the determination of the event centrality.
It consists of two arrays called V0-A and V0-C, each built from 32 scintillator counters and providing full azimuthal coverage, positioned on each side of the interaction point. The V0-A is situated at $z = 3.4$ m ($2.8 < \eta < 5.1$) and the V0-C is located at $z =- 0.9$ m ($-3.7 < \eta < -$ 1.7). Each V0 counter provides the signal amplitude and timing information with a time resolution better than 1 ns~\cite{Aamodt:2008zz, Abbas:2013taa}.
Two Zero Degree Calorimeters (ZDCs)~\cite{Aamodt:2008zz} were used in the offline event selection. The ZDCs are a pair of hadronic calorimeters, one for detecting non-interacting neutrons (ZN) and one for spectator protons (ZP), located at 112.5 m on either side of the interaction point.


\section{Event and track selection}
\label{sec:evTrack}

The data samples analyzed in this article were recorded by ALICE during the 2010 Pb--Pb and 2013 p--Pb runs of the LHC at centre-of-mass energies of $\snn=2.76$ TeV and $\snn=5.02$ TeV, respectively. The Pb--Pb run had equal beam energies, while the p--Pb run had beam energies of 4\,TeV for protons and 1.58\,TeV per nucleon for lead nuclei, which resulted in a rapidity shift of $-0.465$ of the centre-of-mass system with respect to the ALICE laboratory system. In the following, all kinematic variables are reported in the laboratory system. Minimum bias Pb--Pb and p--Pb events were triggered by the coincidence of signals in both V0 detectors. The trigger efficiency is 99.7\% for non-diffractive Pb--Pb collisions~\cite{Abelev:2013qoq} and 99.2\% for non-single-diffractive p--Pb collisions~\cite{ALICE:2012xs}. 
Beam background events were rejected in an offline event selection for all data samples using the timing information from the V0 and ZDC detectors and by correlating reconstructed SPD clusters and tracklets. The remaining beam background was found to be smaller than 0.1\% and was neglected. More details about the offline event selection can be found in~\cite{Abelev:2014ffa}. The fraction of pile-up events in the data sample is found to be negligible after applying dedicated pile-up removal criteria~\cite{Abelev:2014ffa}. Only events with a reconstructed primary vertex within $|z_{vtx} | < 10$ cm with respect to the nominal interaction point were selected. The position of the primary vertex was estimated using tracks reconstructed by the ITS and TPC. The Pb--Pb collision centrality was determined from the measured V0 amplitude distribution~\cite{Abelev:2013qoq}. The dataset of p--Pb collisions is divided into several multiplicity classes defined as fractions of the analysed event sample, based on the charge deposition in the V0-A detector. These multiplicity classes are denoted as ``0--20\%'', ``20--40\%'', ``40--60\%'', and ``60--100\%'', from the highest to the lowest multiplicity. About 13 million Pb--Pb and 92 million p--Pb minimum bias events passed all event selection criteria.  

This analysis used tracks that were reconstructed based on the combined information from the TPC and ITS detectors.
Primary charged tracks were required to have a distance of closest approach to the primary vertex in the longitudinal ($z$) direction and transverse ($xy$) plane smaller than 3.2\,cm and 2.4\,cm, respectively. Tracks with 0.2 $< \pt <$ 5.0~GeV/$c$ were selected in the pseudorapidity range $|\eta|<0.8$, in order to exclude non-uniformities due to the detector boundaries. 
Additional track quality cuts were applied to remove secondary particles (i.e. particles originating from weak decays, photon conversions and secondary hadronic interactions in the detector material) while maintaining good track reconstruction efficiency. 
Tracks were required to have at least 70 TPC space points out of the maximum of 159. 
The $\chi^{2}$ of the track fit per degree of freedom in the TPC reconstruction was required to be below 2.


\section{Systematic uncertainties}
\label{sec:sys} 

The evaluation of systematic uncertainties was performed by varying the event and track selection cuts and by studying the detector response with Monte Carlo (MC) simulations. 
For Pb--Pb, the track selection criteria were changed to only require tracks reconstructed in the TPC alone. This led to a significant difference in most of the observables (up to 10\,\%), which was taken into account in the estimation of the systematic uncertainties. 
Altering the number of TPC space points from 70 to 80, 90 and 100 resulted in a maximum 0.5\% variation of $v_n$ results. 
The variation of the $v_n$ results when using other detectors, e.g. the SPD or TPC, to determine the centrality, is less than 0.5\%. 
No significant variation of the $v_n$ results was seen when altering the polarity of the magnetic field of the ALICE detector, or when narrowing the nominal $|z_{vtx}|$ range from 10 cm to $|z_{vtx}| <$ 7, 8, and 9 cm. The contribution from pileup events to the final systematic uncertainty was found to be negligible. Systematic uncertainties due to detector inefficiencies were investigated using HIJING~\cite{Gyulassy:1994ew} and AMPT~\cite{Lin:2004en} MC simulations. The calculations for a sample at the event generator level (i.e. without invoking either the detector geometry or the reconstruction algorithm) were compared with the results of the analysis of the output of the full reconstruction with a GEANT3~\cite{Brun:1994aa} detector model, in a procedure referred to as an MC closure test.
A difference of up to 4\% for $v_n$ is observed, which is included in the final systematic uncertainty. 
Most of the systematic uncertainties described above cancelled out for $v_{n}\{2\}/v_{n}[2]$ and $r_n$ as indicated in Table~\ref{tab:sysPbPbratio}.
 
For p--Pb collisions, the approach used to evaluate the systematic uncertainty is similar. 
Different track quality cuts are applied, including varying the number of TPC space points, and using tracks reconstructed with the required TPC detector only instead of combined information from TPC and ITS. This leads to a systematic uncertainty of up to 6\% depending on the multiplicity and $\pt$ range.
It was also found that varying the event selection, which includes the cut on the $|z_{vtx}|$, and the cuts to reject pileup events, yields negligible contributions to the final systematic uncertainty. The analysis was repeated using the energy deposited in the neutron ZDC (ZNA) which is located at 112.5 m from the interaction point, instead of using V0-A for the event classes determination. The observed differences with respect to the one using V0-A for event class determination is not included as systematic uncertainty, following the previous paper~\cite{ABELEV:2013wsa}.
In addition, the MC closure is investigated with DPMJET simulations~\cite{Roesler:2000he} combined with GEANT3; this leads to a systematic uncertainty of less than 9\% for $\pt <$ 0.8 GeV/$c$ and 2\% for higher $\pt$. 

The dominant sources of systematic uncertainty are summarized in Tables~\ref{tab:sysPbPbTotal}, \ref{tab:sysPbPbratio} and \ref{tab:syspPbTotal}. The systematic uncertainties evaluated for each of the sources mentioned above were added in quadrature to obtain the total systematic uncertainty of the measurements.

\begin{table}[htb]
  \begin{center}
    \footnotesize
    \begin{tabular*}{150mm}{@{\extracolsep{\fill}}c|cccccccccccc}
        \hline                                                                              
 	Pb--Pb sources  &  $v_{2}\{2\}$  &  $v_{2}[2]$  &  $v_{3}\{2\}$  &  $v_{3}[2]$  & $v_{4}\{2\}$  &  $v_{4}[2]$  \\
       \hline
      Track type & $<$ 4\% & $<$ 4\% &  $<$ 10\%  &  $<$ 8\%  & $<$ 8\%  &  $<$ 8\%    \\
         MC closure 	& $<$ 4\% & $<$ 4\% & $<$ 4\% & $<$ 4\% & $<$ 4\% & $<$ 4\%  \\
           \hline
            Total 	& $<$ 5.7\% &  $<$ 5.7\% & $<$ 10.7\% &  $<$ 9\% &  $<$ 9\%  & $<$ 9\%  \\
         \hline
    \end{tabular*}
    \caption{\label{tab:sysPbPbTotal} Summary of systematic uncertainties of $v_{n}$ for Pb--Pb collisions.}    
  \end{center}
\end{table}

\begin{table}[htb]
  \begin{center}
    \footnotesize
    \begin{tabular*}{150mm}{@{\extracolsep{\fill}}c|ccccc}
        \hline                                                                              
	Pb--Pb sources  &  $v_{2}\{2\} / v_{2}[2]$  &  $v_{3}\{2\} / v_{3}[2]$  & $v_{4}\{2\} / v_{4}[2]$ & $r_{2}$ & $r_{3}$ \\
       \hline
      Track type  & -- & -- & -- &  $<$ 2\%  &  $<$ 5\%  \\
      MC closure 	&  $<$ 1\% &  $<$ 1\% &  $<$ 1\% & $<$ 1\% & $<$ 1\% \\
         \hline
          Total 	& $<$ 1\%  &  $<$ 1\%  & $<$ 1\%  &  $<$ 2.2\%  &  $<$ 5.1\%    \\
         \hline
    \end{tabular*}
    \caption{\label{tab:sysPbPbratio} Summary of systematic uncertainties of $v_{n}\{2\} / v_{n}[2]$ and $r_{n}$ for Pb--Pb collisions.}    
  \end{center}
\end{table}

\begin{table}[htb]
  \begin{center}
    \footnotesize
    \begin{tabular*}{115mm}{@{\extracolsep{\fill}}c|cccccccc}
        \hline
      p--Pb sources  &  $v_{2}\{2\}$  &  $v_{2}[2]$  &  $v_{3}\{2\}$  &  $v_{3}[2]$ &  $v_{2}\{2\} / v_{2}[2]$  & $r_{2}$ &    \\
    \hline
      Track type & $<6\%$ & $<1\%$ & -- & -- & $< 1\%$ & $< 1\%$  \\
      MC closure 	& $<9\%$ & $<8\%$ & $<3\%$ & $<2\%$ & --  & $< 1\%$ \\
        \hline
       Total 	&  $< 10.8\%$ & $<8.1\%$  & $<3\%$ & $<2\%$  &  $<1\%$  & $<1.4\%$ \\
         \hline
    \end{tabular*}
    \caption{\label{tab:syspPbTotal} Summary of systematic uncertainties for p--Pb collisions. }    
  \end{center}
  \end{table}

\section{Results and discussion}
{\label{sec:Results}
\subsection{Pb--Pb collisions}


\begin{figure}[tb]
\begin{center}
\includegraphics[width=0.9\textwidth]{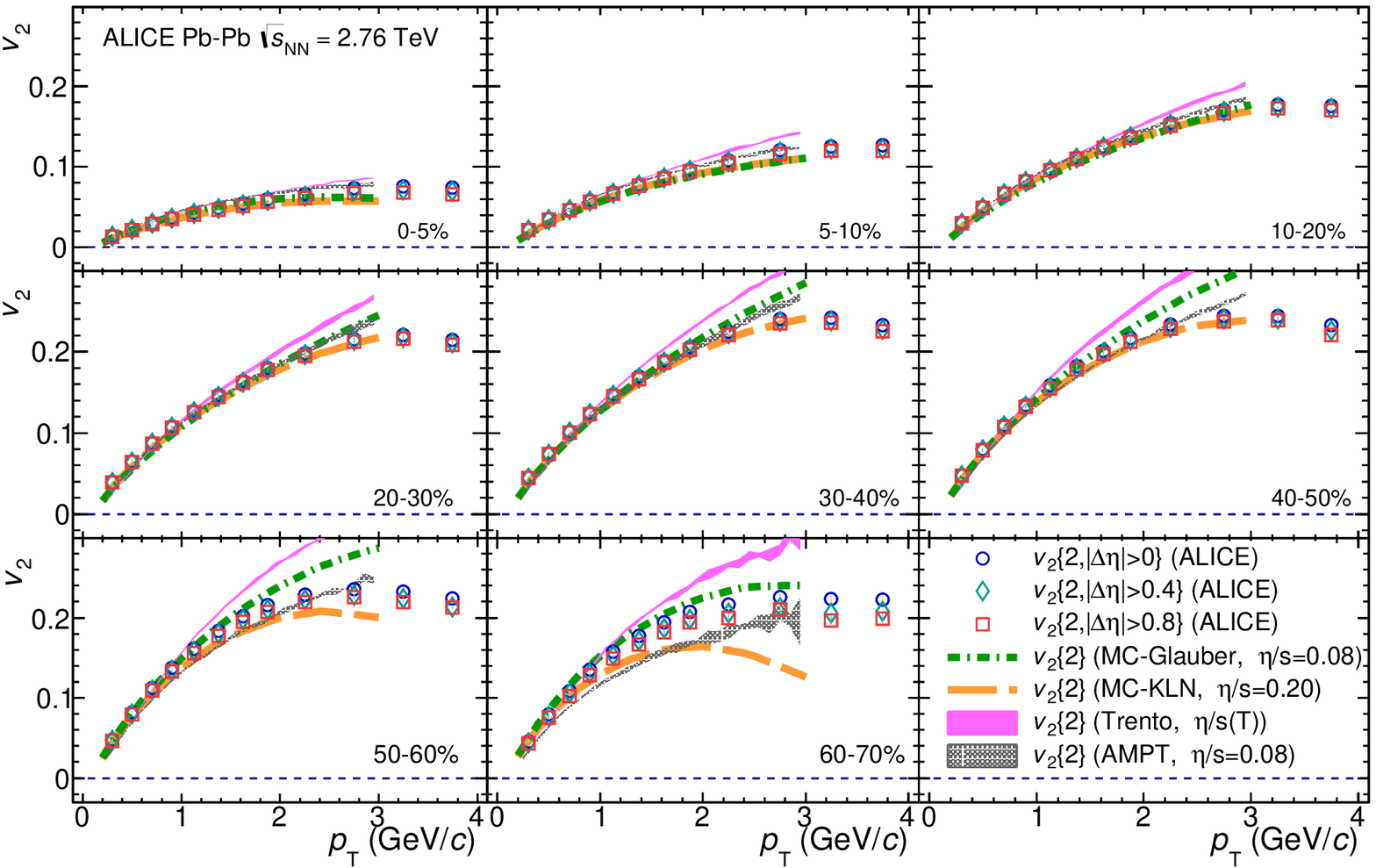}
\caption{$v_{2}\{2\}$ with $|\Delta\eta|>$ 0 (circles), $|\Delta\eta| >$ 0.4 (diamonds) and $|\Delta\eta| >$ 0.8 (squares) for various centrality classes in Pb--Pb collisions at $\sqrt{s_{_{\rm NN}}} = 2.76$ TeV. Hydrodynamic calculations with MC-Glauber initial conditions and $\eta/s =$ 0.08~\cite{Heinz:2013bua}, with MC-KLN initial conditions and $\eta/s =$ 0.20~\cite{Heinz:2013bua}, with Trento initial conditions and temperature dependent $\eta/s$~\cite{Zhao:2017yhj} and AMPT initial conditions and $\eta/s =$ 0.08~\cite{Zhao:2017yhj} are shown in green dot-dash, orange dashed curves, and magenta and grey shaded areas, respectively.}
\label{fig:etagapdep_v22} 
\end{center}
\end{figure}
 
\begin{figure}[tb]
\begin{center}
\includegraphics[width=0.9\textwidth]{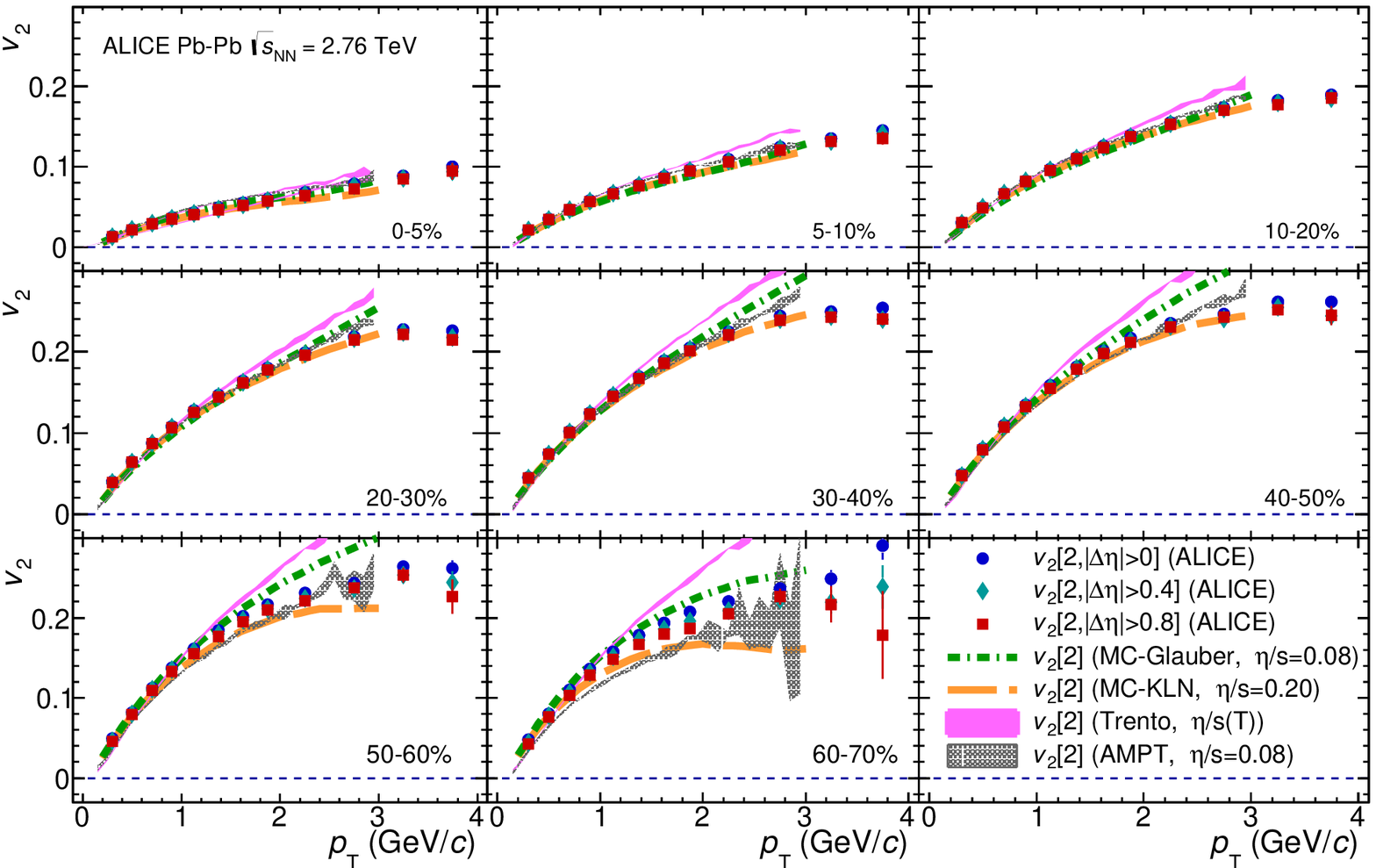}
\caption{$v_{2}[2]$ with $|\Delta\eta|>$ 0 (circles), $|\Delta\eta| >$ 0.4 (diamonds) and $|\Delta\eta| >$ 0.8 (squares) for various centrality classes in Pb--Pb collisions at $\sqrt{s_{_{\rm NN}}} = 2.76$~TeV. Hydrodynamic calculations with MC-Glauber initial conditions~\cite{Heinz:2013bua} and $\eta/s =$ 0.08, with MC-KLN initial conditions and $\eta/s =$ 0.20~\cite{Heinz:2013bua}, with Trento initial conditions and temperature dependent $\eta/s$~\cite{Zhao:2017yhj} and AMPT initial conditions and $\eta/s =$ 0.08~\cite{Zhao:2017yhj} are shown in green dot-dashed and orange dashed curves, and magenta and grey shaded areas, respectively.}
\label{fig:etagapdep_newv22} 
\end{center}
\end{figure}

\begin{figure}[tb]
\begin{center}
\includegraphics[width=0.9\textwidth]{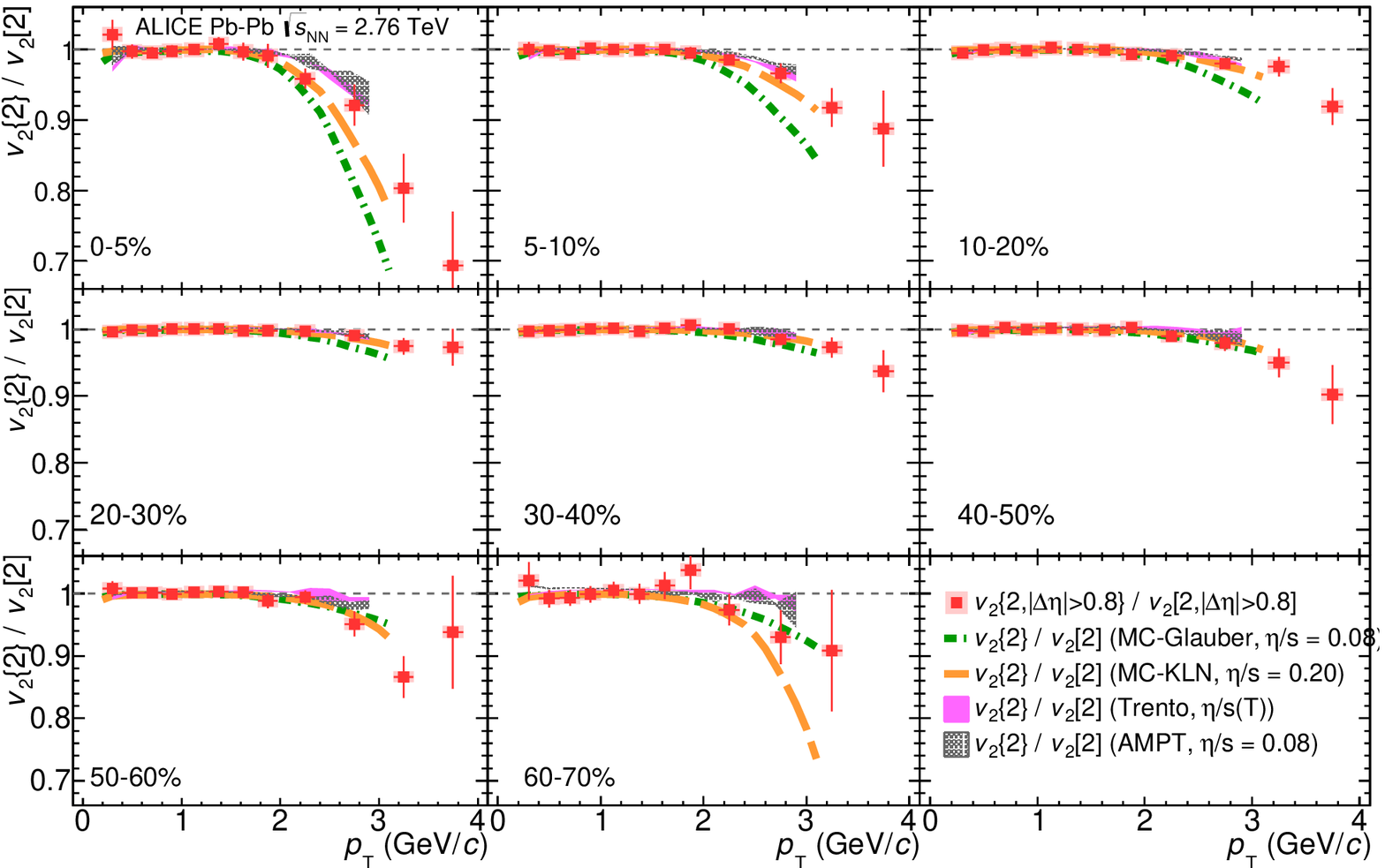}
\caption{The ratio $v_{2}\{2, |\Delta\eta| > 0.8\} / v_{2}[2, |\Delta\eta| > 0.8]$ in Pb--Pb collisions at $\sqrt{s_{_{\rm NN}}} = 2.76$~TeV. The different panels show the centrality evolution of the measurements.
Hydrodynamic calculations with MC-Glauber initial conditions and $\eta/s =$ 0.08~\cite{Heinz:2013bua}, with MC-KLN initial conditions and $\eta/s =$ 0.20~\cite{Heinz:2013bua}, with Trento initial conditions and temperature dependent $\eta/s$~\cite{Zhao:2017yhj} and AMPT initial conditions and $\eta/s =$ 0.08~\cite{Zhao:2017yhj} are shown in green dot-dashed and orange dashed curves, and magenta and grey shaded areas, respectively.}
\label{fig:ratio_newv22overv22} 
\end{center}
\end{figure}

Figures~\ref{fig:etagapdep_v22} and~\ref{fig:etagapdep_newv22} show the $\pt$ dependence of $v_{2}\{2\}$ and $v_{2}[2]$ with three different pseudorapidity gaps, for centrality classes from 0--5\% to 70--80\%. 
The analysed events are divided into two sub-events A and B, separated by a pseudorapidity gap. Note that $|\Delta\eta|>$ 0 suggests that there is no separation in pseudorapidity between the two sub-events. Short-range correlations, one of the main sources of non-flow effects, are expected to be suppressed when using a large pseudorapidity gap. It is observed that $v_{2}\{2\}$ and $v_{2}[2]$ using various pseudorapidity gaps do not change significantly for central and semi-central collisions. The decrease of $v_{2}$ with larger pseudorapidity gaps is more prominent in the most peripheral collisions, mainly due to the suppression of non-flow effects. The results are also compared to the original predictions within the $\tt VISH2$+$\tt 1$ hydrodynamic framework with: 1) Monte Carlo Glauber (MC-Glauber) initial conditions and $\eta/s =$ 0.08; 2) Monte Carlo Kharzeev-Levin-Nardi (MC-KLN) initial conditions and $\eta/s =$ 0.20~\cite{Heinz:2013bua}. In addition, the comparisons to recently released calculations from the $\tt iEBE$-$\tt VISHNU$ hydrodynamic framework with: 1) Trento initial conditions, temperature dependent shear and bulk viscosities, $\eta/s({\rm T})$ and $\zeta({\rm T})$; and 2) AMPT initial conditions with  $\eta/s =$ 0.08~\cite{Zhao:2017yhj} are also presented. These combinations of various initial conditions and $\eta/s$ are chosen due to the fact that they give the best descriptions of the particle spectra and the integrated flow measurements~\cite{Qiu:2011hf,Zhao:2017yhj}. The four hydrodynamic calculations describe the $v_2\{2\}$ very well up to $p_{\rm T} \approx$ 2 GeV/$c$ at least for central and semi-central collisions, as do the calculations with MC-Glauber, MC-KLN and AMPT initial conditions for the $v_2[2]$. For central and mid-central collisions, calculations with MC-KLN and AMPT initial conditions predict both $v_2\{2\}$ and $v_2[2]$ better for higher $p_{\rm T}$ than those with MC-Glauber and Trento initial conditions. For more peripheral collisions, the experimental $v_2$ data in both cases fall between the four sets of predictions.

In order to probe the $p_{\rm T}$-dependent flow vector fluctuations quantitatively, the ratio $v_{2}\{2, |\Delta\eta| > 0.8\}$ $/v_{2}[2, |\Delta\eta| > 0.8]$ using Eq.~\ref{eq:v22EQ} is presented as a function of $p_{\rm T}$ for different centrality classes in Fig.~\ref{fig:ratio_newv22overv22}. This ratio is consistent with unity up to $p_{\rm T} \approx$ 2 GeV/$c$ and starts to deviate from unity in the higher $\pt$ region in the most central collisions. The deviations from unity are weak and within 10$\%$ in non-central collisions in the presented $p_{\rm T}$ range. To better understand whether such deviations from unity are caused by non-flow effects, the like-sign technique, which suppresses contributions from resonance decays by correlating only particles with same charge, is applied. The differences of the measured $v_{2}\{2, |\Delta\eta| > 0.8\}/v_{2}[2, |\Delta\eta| > 0.8]$ from like-sign and all charged particles are found to be less than 0.5\%. This shows that deviations of $v_{2}\{2, |\Delta\eta| > 0.8\}/v_{2}[2, |\Delta\eta| > 0.8]$ from unity cannot be explained solely by non-flow effects from resonance decays. It is also seen in Fig.~\ref{fig:ratio_newv22overv22} that the hydrodynamic calculations with MC-KLN, Trento and AMPT initial conditions describe the data fairly well for all centrality classes except for the most peripheral collisions, while MC-Glauber calculations reproduce the data only for mid-central and peripheral collisions. This indicates that hydrodynamic calculations with AMPT and MC-KLN initial conditions and $\eta/s =$ 0.20 not only generate reasonable $v_{2}$ values, but also reproduce the measured $v_{2}\{2, |\Delta\eta| > 0.8\}/v_{2}[2, |\Delta\eta| > 0.8]$.

\begin{figure}
\begin{center}
\includegraphics[width=0.9\textwidth]{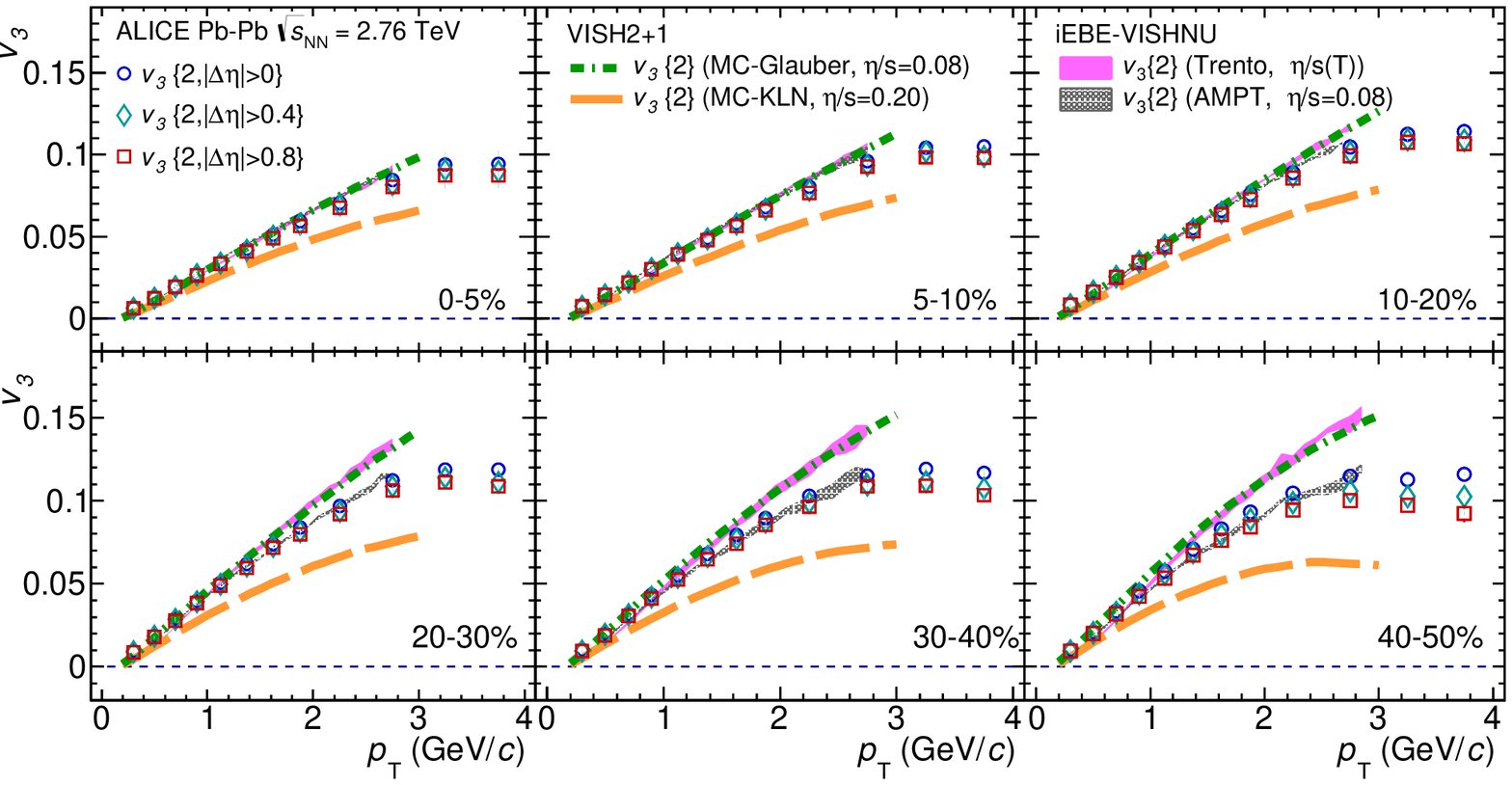}
\caption{$v_{3}\{2\}$ with different $|\Delta\eta|$ gaps is presented in Pb--Pb collisions at $\sqrt{s_{_{\rm NN}}} = 2.76$~TeV. $v_{3}\{2, |\Delta\eta|> 0\}$, $v_{3}\{2, |\Delta\eta|> 0.4\}$, and $v_{3}\{2, |\Delta\eta|> 0.8\}$ are represented by circles, diamonds and squares, respectively. The different panels show the centrality evolution of the measurements.
Hydrodynamic calculations with MC-Glauber initial conditions and $\eta/s =$ 0.08~\cite{Heinz:2013bua}, with MC-KLN initial conditions and $\eta/s =$ 0.20~\cite{Heinz:2013bua}, with Trento initial conditions and temperature dependent $\eta/s$~\cite{Zhao:2017yhj} and AMPT initial conditions and $\eta/s =$ 0.08~\cite{Zhao:2017yhj} are shown in green dot-dash, orange dashed curves, and magenta and grey shaded areas, respectively.}
\label{fig:etagapdep_v32} 
\end{center}
\end{figure}

\begin{figure}
\begin{center}
\includegraphics[width=0.9\textwidth]{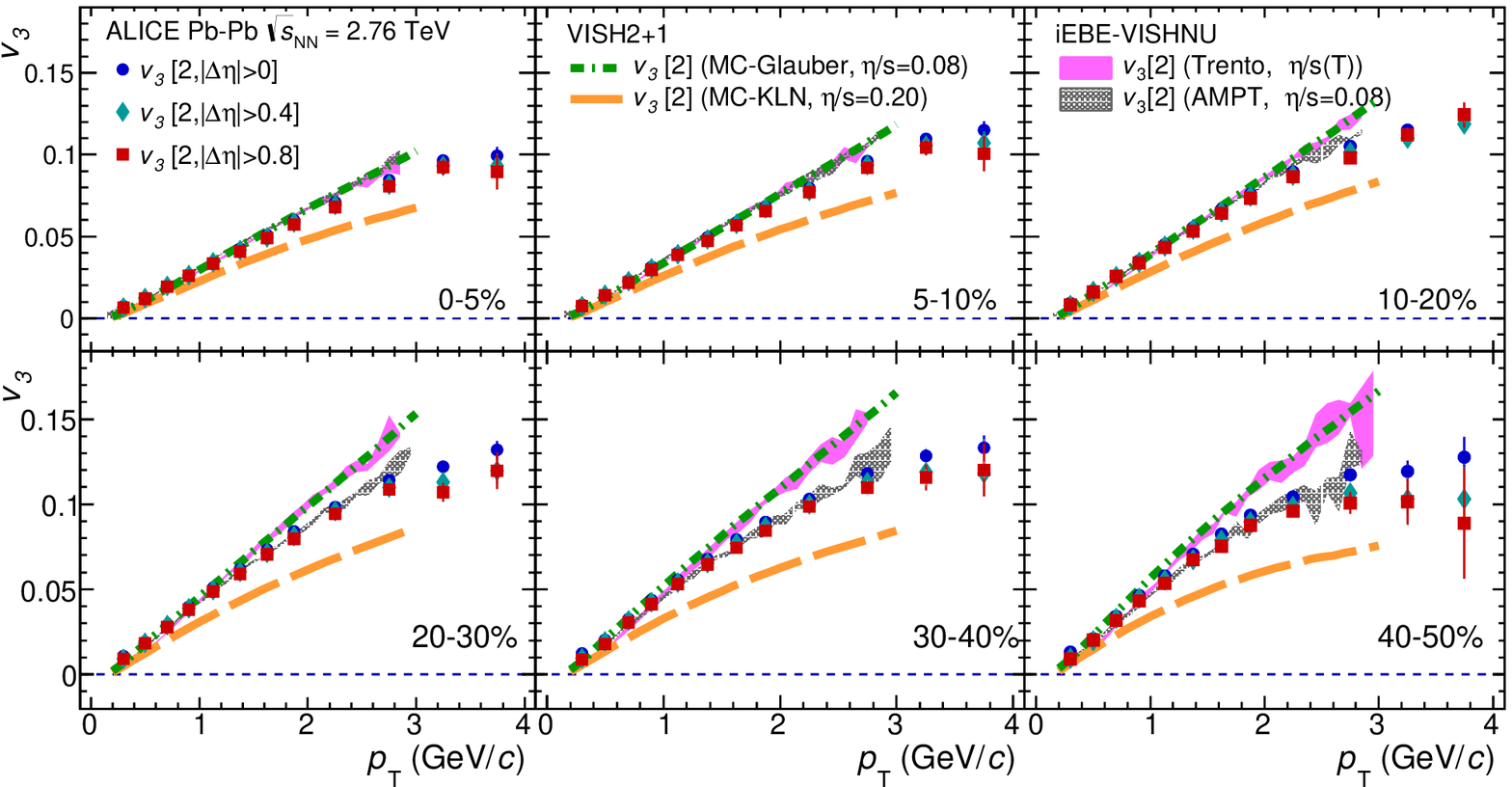}
\caption{$v_{3}[2]$ with different $|\Delta\eta|$ gaps is presented in Pb--Pb collisions at $\sqrt{s_{_{\rm NN}}} = 2.76$~TeV. $v_{3}[2, |\Delta\eta|> 0]$, $v_{3}[2, |\Delta\eta|> 0.4]$, and $v_{3}[2, |\Delta\eta|> 0.8]$ are represented by circles, diamonds, and squares, respectively. The different panels show the centrality evolution of the measurements.
Hydrodynamic calculations with MC-Glauber initial conditions and $\eta/s =$ 0.08~\cite{Heinz:2013bua} and with MC-KLN initial conditions and $\eta/s =$ 0.20~\cite{Heinz:2013bua}, with Trento initial conditions and temperature dependent $\eta/s$~\cite{Zhao:2017yhj} and AMPT initial conditions and $\eta/s =$ 0.08~\cite{Zhao:2017yhj} are shown in green dot-dash, orange dashed curves, and magenta and grey shaded areas, respectively.}
\label{fig:etagapdep_newv32} 
\end{center}
\end{figure}

\begin{figure}
\begin{center}
\includegraphics[width=0.9\textwidth]{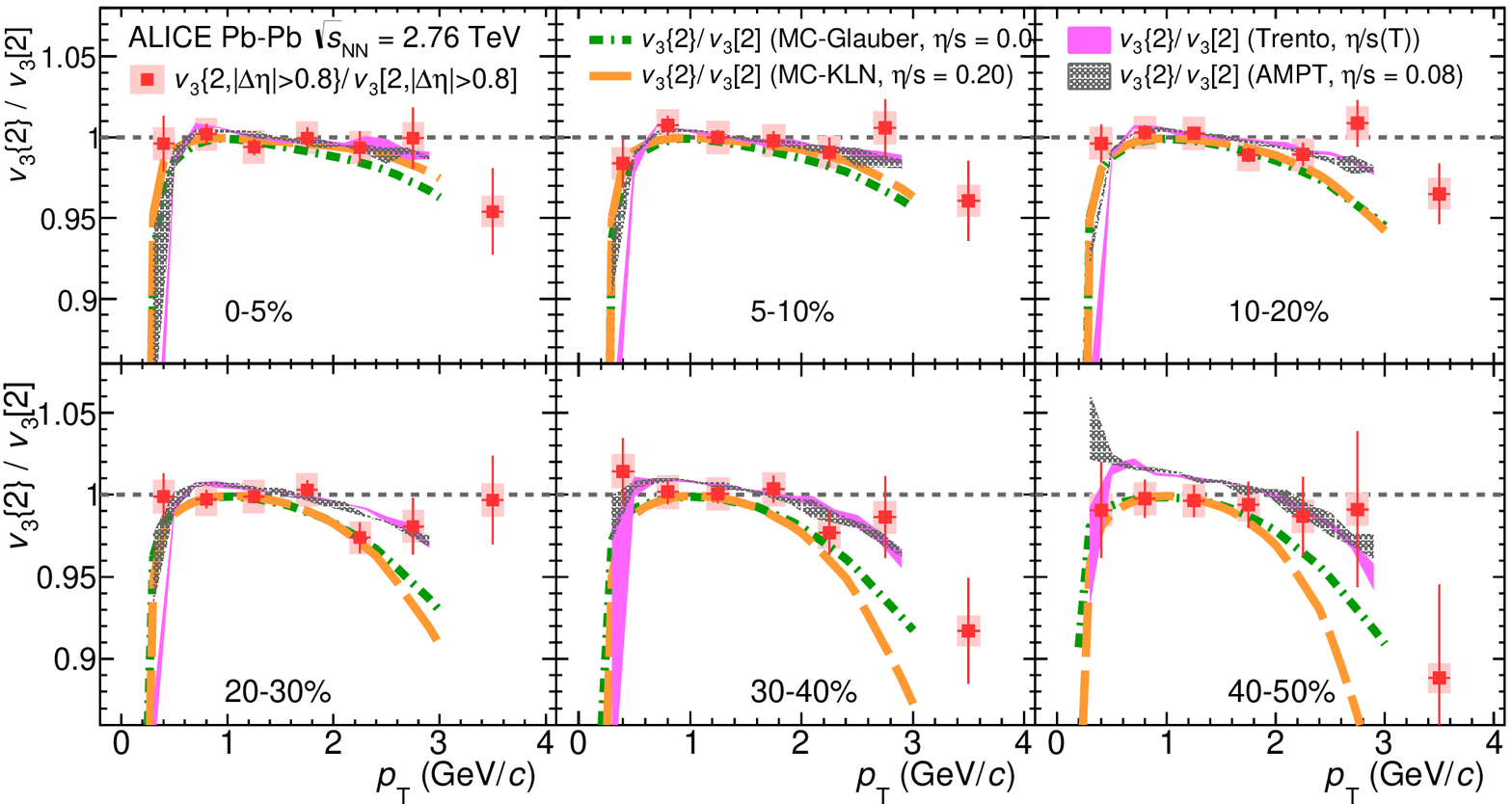}
\caption{The ratio $v_{3}\{2, |\Delta\eta| > 0.8\} / v_{3}[2, |\Delta\eta| > 0.8]$ in Pb--Pb collisions at $\sqrt{s_{_{\rm NN}}} = 2.76$~TeV. The different panels show the centrality evolution of the measurements.
Hydrodynamic calculations with MC-Glauber initial conditions and $\eta/s =$ 0.08~\cite{Heinz:2013bua} and with MC-KLN initial conditions and $\eta/s =$ 0.20~\cite{Heinz:2013bua}, with Trento initial conditions and temperature dependent $\eta/s$~\cite{Zhao:2017yhj} and AMPT initial conditions and $\eta/s =$ 0.08~\cite{Zhao:2017yhj} are shown in green dot-dash, orange dashed curves, and magenta and grey shaded areas, respectively.}
\label{fig:ratio_v3} 
\end{center}
\end{figure}

The higher order anisotropic flow coefficients, which were first measured in~\cite{ALICE:2011ab}, are shown to be more sensitive to the initial conditions and $\eta/s$~\cite{Alver:2010dn}. In Figs.~\ref{fig:etagapdep_v32} and~\ref{fig:etagapdep_newv32}, $v_{3}\{2\}$ and $v_{3}[2]$ are shown with three different pseudorapidity gaps for several centrality classes. Similar to what was presented in Figs.~\ref{fig:etagapdep_v22} and~\ref{fig:etagapdep_newv22}, both $v_{3}\{2\}$ and $v_{3}[2]$ show a decreasing trend as the pseudorapidity gap increases, in particular in more peripheral collisions. Only a weak centrality dependence is observed for both $v_{3}\{2\}$ and $v_{3}[2]$. The comparison to hydrodynamic calculations demonstrates that although hydrodynamic calculations with MC-Glauber and MC-KLN initial conditions roughly describe $v_{2}\{2\}$ and $v_{2}[2]$, they cannot describe $v_{3}\{2\}$ and $v_{3}[2]$ over the full $\pt$ range and for all centrality classes, and tend to overpredict or underpredict the data. Similar as $v_{2}$, the hydrodynamic calculation with Trento initial conditions overestimates both $v_{3}\{2\}$ and $v_{3}[2]$ measurements, while the one with AMPT initial conditions quantitatively describe the measured $v_{3}$ for presented $\pt$ and centrality intervals.

The ratio $v_{3}\{2, |\Delta\eta| > 0.8\}/v_{3}[2, |\Delta\eta| > 0.8]$ is shown together with hydrodynamic calculations in Fig.~\ref{fig:ratio_v3}. Wider $\pt$ intervals were used for the ratio than for the individual $v_3$ measurements in order to suppress statistical fluctuations. It was found that the ratio agrees with unity over a wide $p_{\rm T}$ range, as opposed to $v_{2}\{2, |\Delta\eta| > 0.8\}/v_{2}[2, |\Delta\eta| > 0.8]$. No clear indication of $p_{\rm T}$-dependent $V_{3}$ flow vector fluctuations are observed for the presented centrality and $\pt$ regions within the large uncertainties. Despite the fact that the hydrodynamic calculations with MC-Glauber and MC-KLN initial conditions cannot reproduce the magnitude of $v_{3}\{2\}$ and $v_{3}[2]$, the validity of the two sets of initial conditions could be examined also by the comparison of the predicted $v_{3}\{2\}/v_{3}[2]$ ratio to data, which should be independent of the magnitude of $v_{3}$. Hydrodynamic calculations from $\tt VISH2$+$\tt 1$, especially the one with MC-KLN initial conditions, overestimate the possible $p_{\rm T}$-dependent $V_{3}$ flow vector fluctuations, despite the good description for the second harmonic. A good agreement between data and hydrodynamic calculations from $\tt iEBE$-$\tt VISHNU$ is found for all centrality intervals. This is expected for AMPT initial conditions as the calculations quantitatively reproduce both measured $v_{3}\{2\}$ and $v_{3}[2]$ as discussed above. However, the calculations with Trento initial conditions, which overestimate both $v_{3}\{2\}$ and $v_{3}[2]$, are consistent with the measured $v_{3}\{2, |\Delta\eta| > 0.8\}/v_{3}[2, |\Delta\eta| > 0.8]$ ratio. This accidental agreement needs further investigations in the $\tt iEBE$-$\tt VISHNU$ framework to understand the physics mechanism responsible for this behaviour.


\begin{figure}
\begin{center}
\includegraphics[width=0.9\textwidth]{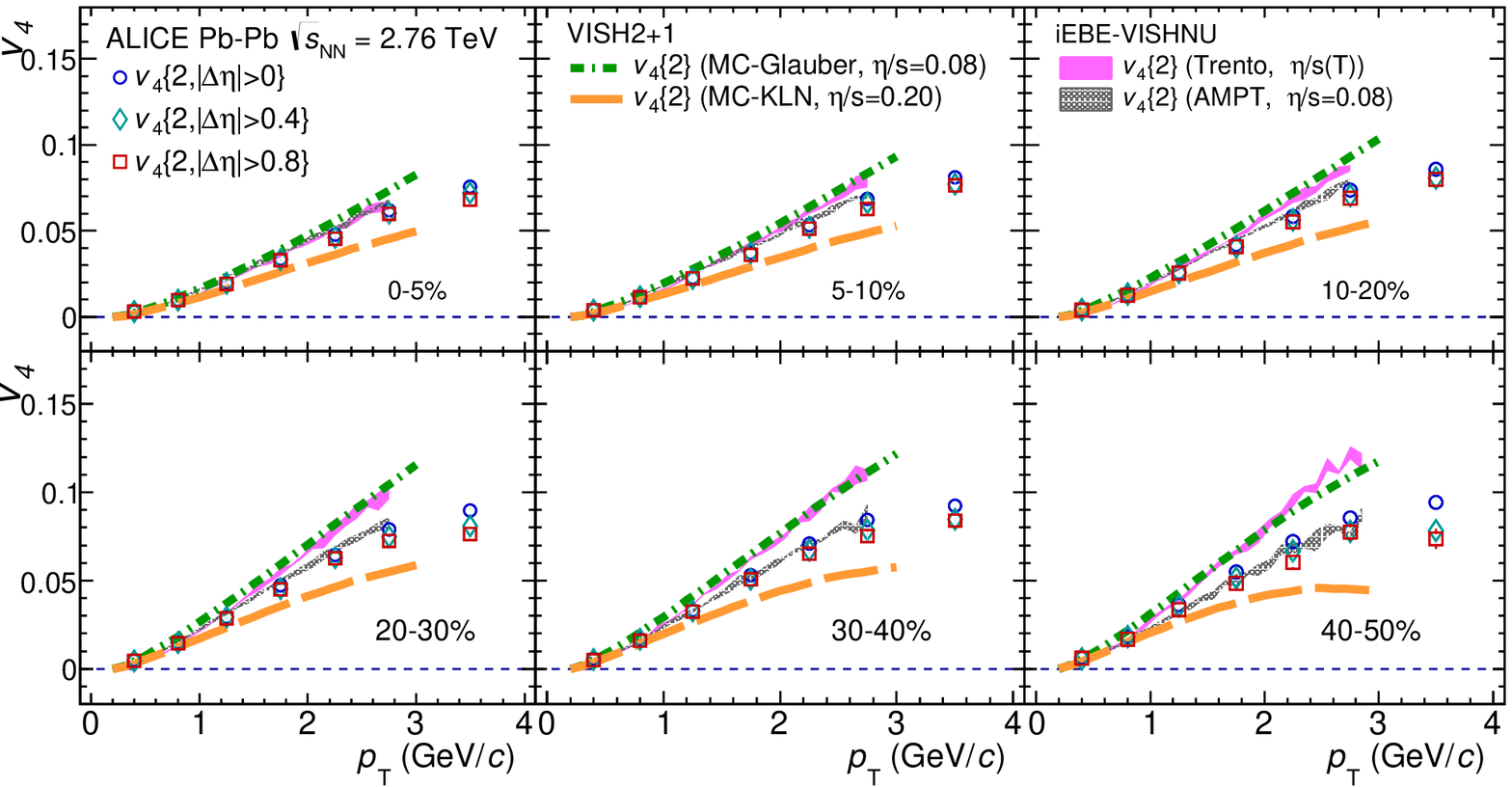}
\caption{$v_{4}\{2\}$ with different $|\Delta\eta|$ gaps is presented in Pb--Pb collisions at $\sqrt{s_{_{\rm NN}}} = 2.76$~TeV. $v_{4}\{2, |\Delta\eta|> 0\}$, $v_{4}\{2, |\Delta\eta|> 0.4\}$ and $v_{4}\{2, |\Delta\eta|> 0.8\}$ are represented by circles, diamonds, and squares, respectively. The different panels show the centrality evolution of the measurements.
Hydrodynamic calculations with MC-Glauber initial conditions and $\eta/s =$ 0.08~\cite{Heinz:2013bua}, with MC-KLN initial conditions and $\eta/s =$ 0.20~\cite{Heinz:2013bua}, with Trento initial conditions and temperature dependent $\eta/s$~\cite{Zhao:2017yhj} and AMPT initial conditions and $\eta/s =$ 0.08~\cite{Zhao:2017yhj} are shown in green dot-dash, orange dashed curves, and magenta and grey shaded areas, respectively.}
\label{fig:v42}
\end{center} 
\end{figure}

\begin{figure}
\begin{center}
\includegraphics[width=0.9\textwidth]{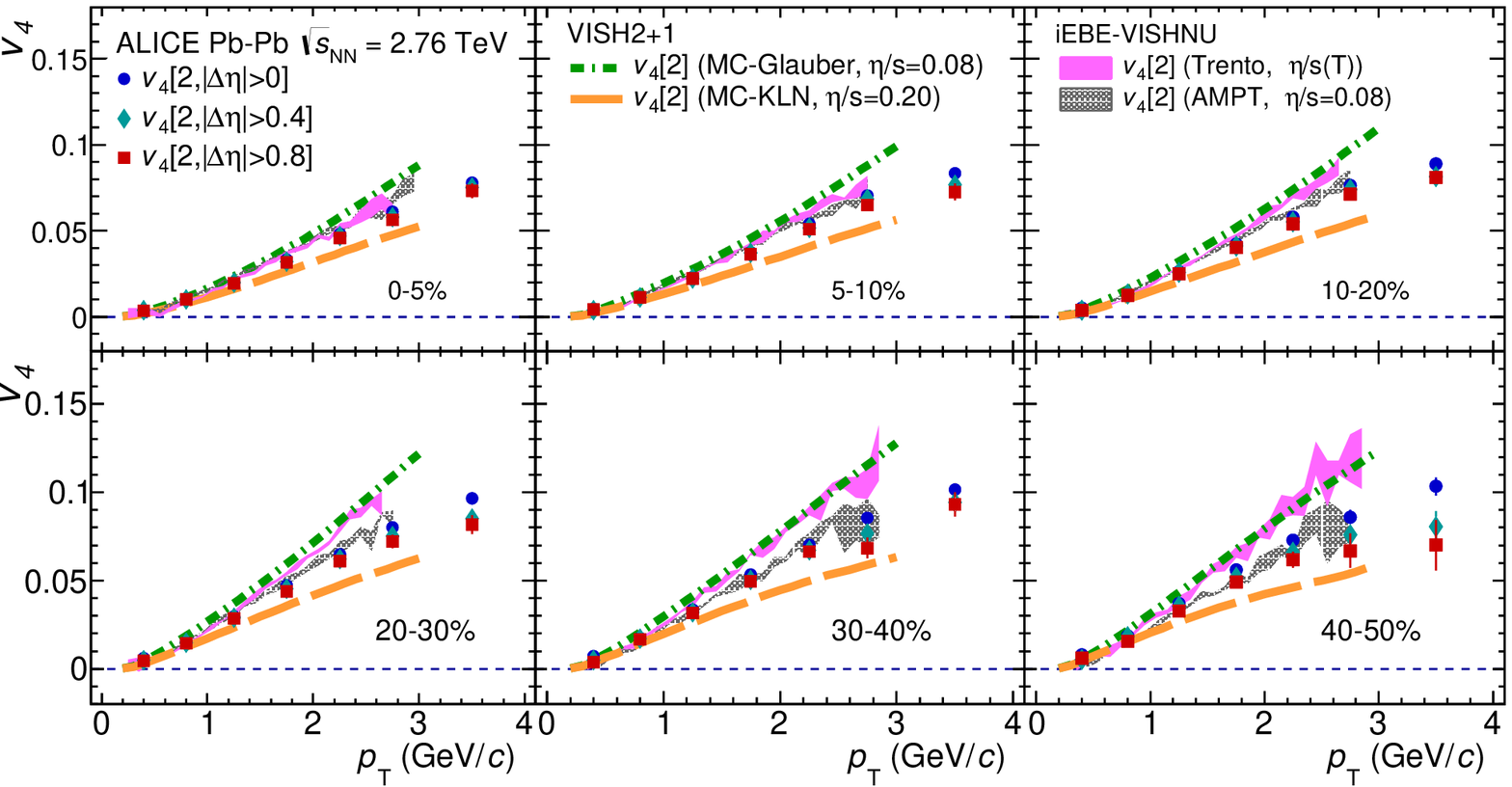}
\caption{$v_{4}[2]$ with different $|\Delta\eta|$ gaps is presented in Pb--Pb collisions at $\sqrt{s_{_{\rm NN}}} = 2.76$~TeV. $v_{4}[2, |\Delta\eta|> 0]$, $v_{4}[2, |\Delta\eta|> 0.4]$, and $v_{4}[2, |\Delta\eta|> 0.8]$ are represented by circles, diamonds, and squares, respectively. The different panels show the centrality evolution of the measurements.
Hydrodynamic calculations with MC-Glauber initial conditions and $\eta/s =$ 0.08~\cite{Heinz:2013bua} and with MC-KLN initial conditions and $\eta/s =$ 0.20~\cite{Heinz:2013bua}, with Trento initial conditions and temperature dependent $\eta/s$~\cite{Zhao:2017yhj} and AMPT initial conditions and $\eta/s =$ 0.08~\cite{Zhao:2017yhj} are shown in green dot-dash, orange dashed curves, and magenta and grey shaded areas, respectively.}
\label{fig:newv42} 
\end{center}
\end{figure}

\begin{figure}
\begin{center}
\includegraphics[width=0.9\textwidth]{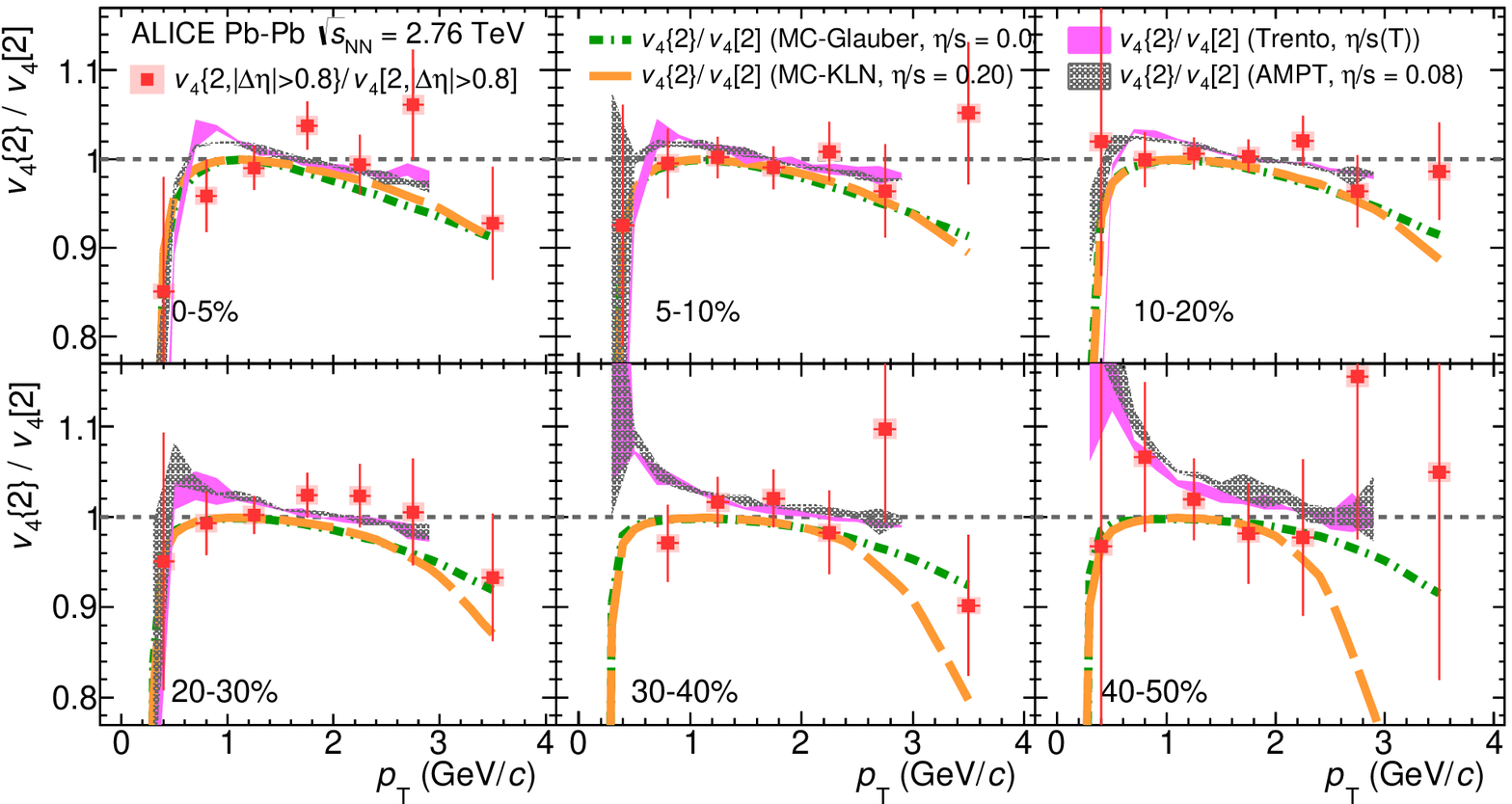}
\caption{The ratio $v_{4}\{2, |\Delta\eta| > 0.8\} / v_{4}[2, |\Delta\eta| > 0.8]$ in Pb--Pb collisions at $\sqrt{s_{_{\rm NN}}} = 2.76$~TeV. The different panels show the centrality evolution of the measurements. Hydrodynamic calculations with MC-Glauber initial conditions and $\eta/s =$ 0.08~\cite{Heinz:2013bua} and with MC-KLN initial conditions and $\eta/s =$ 0.20~\cite{Heinz:2013bua}, with Trento initial conditions and temperature dependent $\eta/s$~\cite{Zhao:2017yhj} and AMPT initial conditions and $\eta/s =$ 0.08~\cite{Zhao:2017yhj} are shown in green dot-dash, orange dashed curves, and magenta and grey shaded areas, respectively.}
\label{fig:ratiov42} 
\end{center}
\end{figure}

\begin{figure}
\begin{center}
\includegraphics[width=0.75\textwidth]{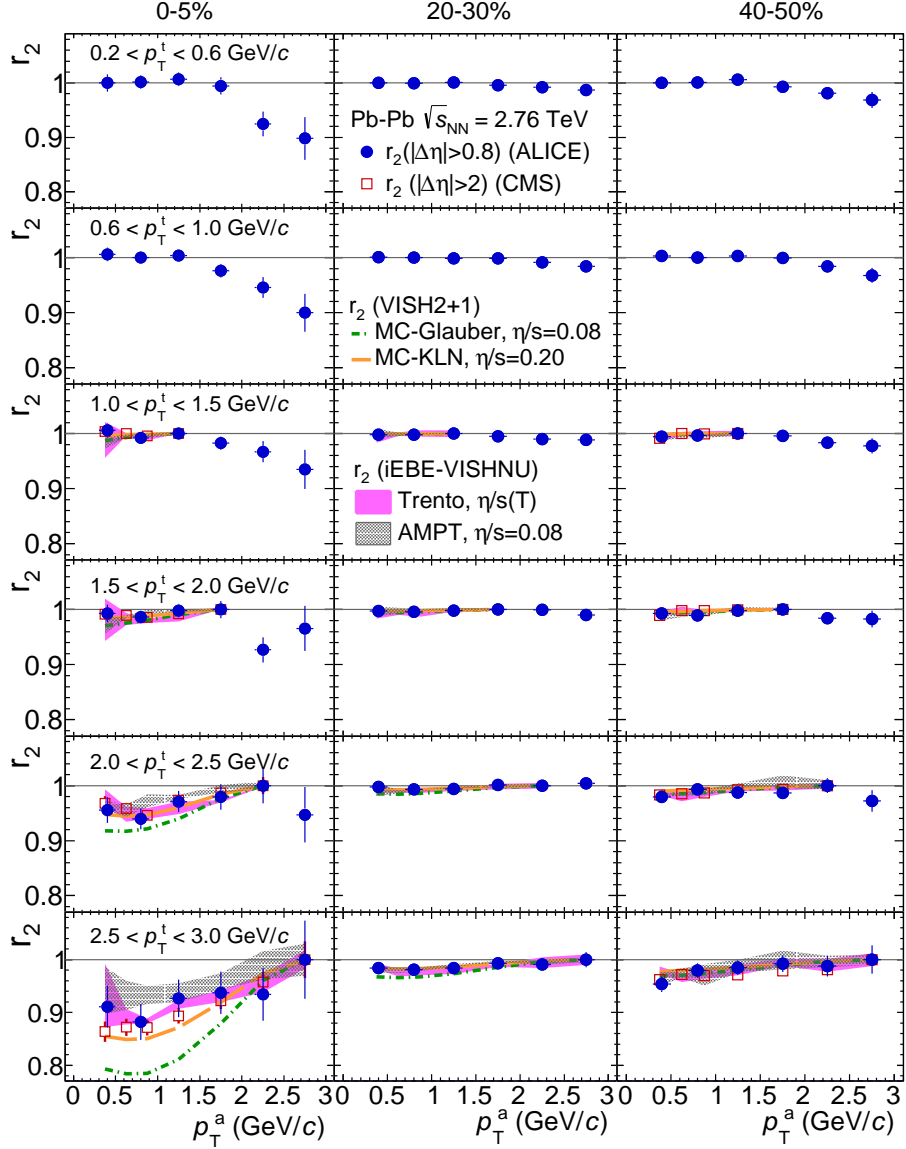}
\caption{The factorisation ratio $r_{2}$, as a function of $p_{\rm T}^{\, a}$ in bins of $p_{\rm T}^{t}$ for 0--5$\%$, 20--30$\%$ and 40--50$\%$ centralities in Pb--Pb collisions at $\sqrt{s_{_{\rm NN}}} = 2.76$~TeV, is presented (solid circles). CMS measurements are presented by open square~\cite{CMS:2013bza}. Hydrodynamic calculations with MC-Glauber initial conditions and $\eta/s =$ 0.08~\cite{Heinz:2013bua} and with MC-KLN initial conditions and $\eta/s =$ 0.20~\cite{Heinz:2013bua}, with Trento initial conditions and temperature dependent $\eta/s$~\cite{Zhao:2017yhj} and AMPT initial conditions and $\eta/s =$ 0.08~\cite{Zhao:2017yhj} are shown in green dot-dash, orange dashed curves, and magenta and grey shaded areas, respectively.}
\label{fig:r2etagaps} 
\end{center}
\end{figure}

The centrality dependence of $v_{4}\{2\}$ and $v_{4}[2]$ with three different pseudorapidity gaps are shown in Figs.~\ref{fig:v42} and~\ref{fig:newv42}. Decreasing trends with increasing $|\Delta \eta|$ gaps and a weak centrality dependence are observed for both measurements. The hydrodynamic calculations with MC-Glauber and Trento initial conditions overestimate the measurements of $v_{4}\{2\}$ and $v_{4}[2]$, while the calculations with MC-KLN initial conditions underestimate the measurements, similar to what was seen for the $v_3$ observables. On the other hand, the hydrodynamic calculations from AMPT initial conditions agree with the measurements of $v_{4}\{2\}$ and $v_{4}[2]$. Moreover, the ratio $v_{4}\{2, |\Delta\eta| > 0.8\}/v_{4}[2, |\Delta\eta| > 0.8]$ shown in Fig.~\ref{fig:ratiov42} is in agreement with unity albeit with large uncertainties for the presented $p_{\rm T}$ range and centrality classes. The validity of the hydrodynamic calculations cannot be judged due to the large uncertainties of the $v_{4}\{2, |\Delta\eta| > 0.8\}/v_{4}[2, |\Delta\eta| > 0.8]$ measurements.


Alternatively, one can search for $p_{\rm T}$-dependent flow vector fluctuations via the measurement of the factorisation ratio, $r_{n}$. The results of $r_{2}$ and $r_{3}$ are presented in Figs.~\ref{fig:r2etagaps} and \ref{fig:r3etagaps} as a function of $p_{\rm T}^{t}$ and $p_{\rm T}^{\, a}$ with $|\Delta\eta|>0.8$ for three centrality classes in Pb--Pb collisions at $\sqrt{s_{_{\rm NN}}} = $ 2.76 TeV. By construction, $r_{n} = 1$ when the triggered and associated particles are from the same $p_{\rm T}$ interval. 
In contrast to the previous analysis~\cite{Aamodt:2011by}, there is no $p_{\rm T}^{t} \geq p_{\rm T}^{\, a}$ cut applied here to avoid auto-correlations (taking the same particle as both triggered and associated particles in the two-particle azimuthal correlations). The triggered particles are always selected from the negative pseudorapidity region and the associated particles are from the positive pseudorapidity region.
The $r_{2}$ value deviates significantly from unity for the most central collisions. This effect becomes stronger with an increasing difference between $p_{\rm T}^{t}$ and $p_{\rm T}^{\, a}$. 
The previous results indicated that factorisation holds approximately for $n \geq$ 2 and $p_{\rm T}$ below 4 GeV/$c$, while deviations emerging at higher $\pt$ were ascribed to recoil jet contributions~\cite{Aamodt:2011by}. This analysis, however, shows that factorisation breaks down at lower $p_{\rm T}$ when the more sensitive observable, $r_{2}$, is used.
The deviation reaches 10$\%$ for the lowest $p_{\rm T}^{\, a}$ in the 0--5$\%$ centrality range, for 2.5 $\textless~p_{\rm T}^{t}~\textless$ 3 GeV/$c$. One explanation from~\cite{Gardim:2012im} is that this deviation is due to the $p_{\rm T}$-dependent $V_{2}$ flow vector fluctuations, which originate from initial event-by-event geometry fluctuations. 
Hydrodynamic calculations~\cite{Heinz:2013bua} are compared to data for the presented centrality classes and for selected $\pt$ bins. Both hydrodynamic calculations from $\tt VISH2$+$\tt 1$ and $\tt iEBE$-$\tt VISHNU$ frameworks qualitatively predict the trend of $r_{2}$, while the data agree quantitatively better with $\tt iEBE$-$\tt VISHNU$.
In addition, the CMS measurements~\cite{CMS:2013bza, Khachatryan:2015oea} are consistent with our measurements.

\begin{figure}
\begin{center}
\includegraphics[width=0.75\textwidth]{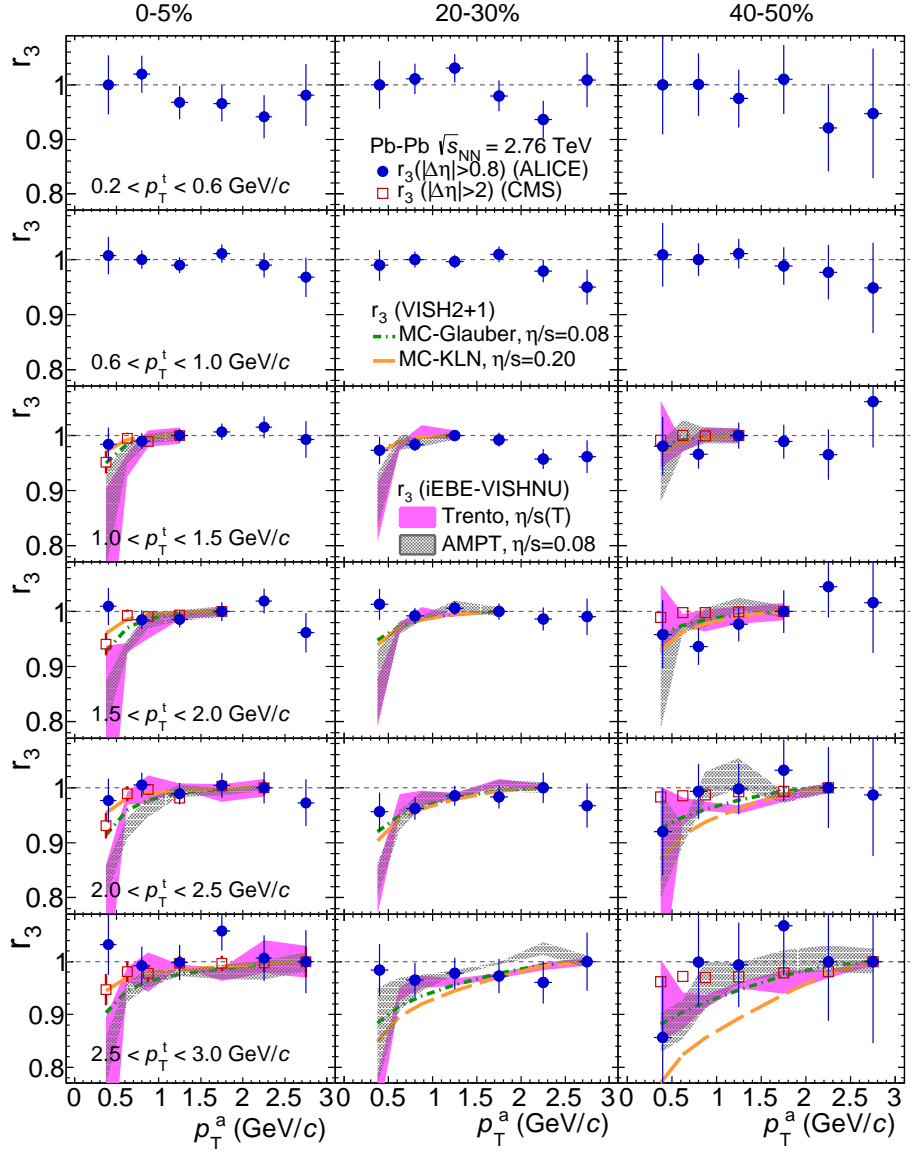}
\caption{The factorisation ratio $r_{3}$, as a function of $p_{\rm T}^{\, a}$ in bins of $p_{\rm T}^{t}$ for 0--5$\%$, 20--30$\%$ and 40--50$\%$ centralities in Pb--Pb collisions at $\sqrt{s_{_{\rm NN}}} = 2.76$~TeV, is presented  (solid circles). CMS measurements~\cite{CMS:2013bza} are presented by open squares. Hydrodynamic calculations with MC-Glauber initial conditions and $\eta/s =$ 0.08~\cite{Heinz:2013bua} and with MC-KLN initial conditions and $\eta/s =$ 0.20~\cite{Heinz:2013bua}, with Trento initial conditions and temperature dependent $\eta/s$~\cite{Zhao:2017yhj} and AMPT initial conditions and $\eta/s =$ 0.08~\cite{Zhao:2017yhj} are shown in green dot-dash, orange dashed curves, and magenta and grey shaded areas, respectively.}
\label{fig:r3etagaps} 
\end{center}
\end{figure}

For $r_{3}$, the results are compatible with unity, and can be described by hydrodynamic calculations from both $\tt VISH2$+$\tt 1$ and $\tt iEBE$-$\tt VISHNU$ frameworks, albeit with large statistical uncertainties. The factorisation is valid over a wider range of $p_{\rm T}^{\, a}$, $p_{\rm T}^{t}$ and centrality ranges, as opposed to $r_{2}$.
The possible breakdown of factorisation, if it exists, is within 10$\%$ when both $p_{\rm T}^{\, a}$ and $p_{\rm T}^{t}$ are below 3 GeV/$c$. 
The CMS measurements~\cite{CMS:2013bza, Khachatryan:2015oea} are consistent with the $r_{3}$ results presented here despite the fact that the pseudorapidity gaps are different between the two measurements. 
Better agreements with hydrodynamic calculations are observed with $\tt VISH2$+$\tt 1$. 

\subsection{p--Pb collisions}

\begin{figure}[h]
\begin{center}
\includegraphics[width=0.9\textwidth]{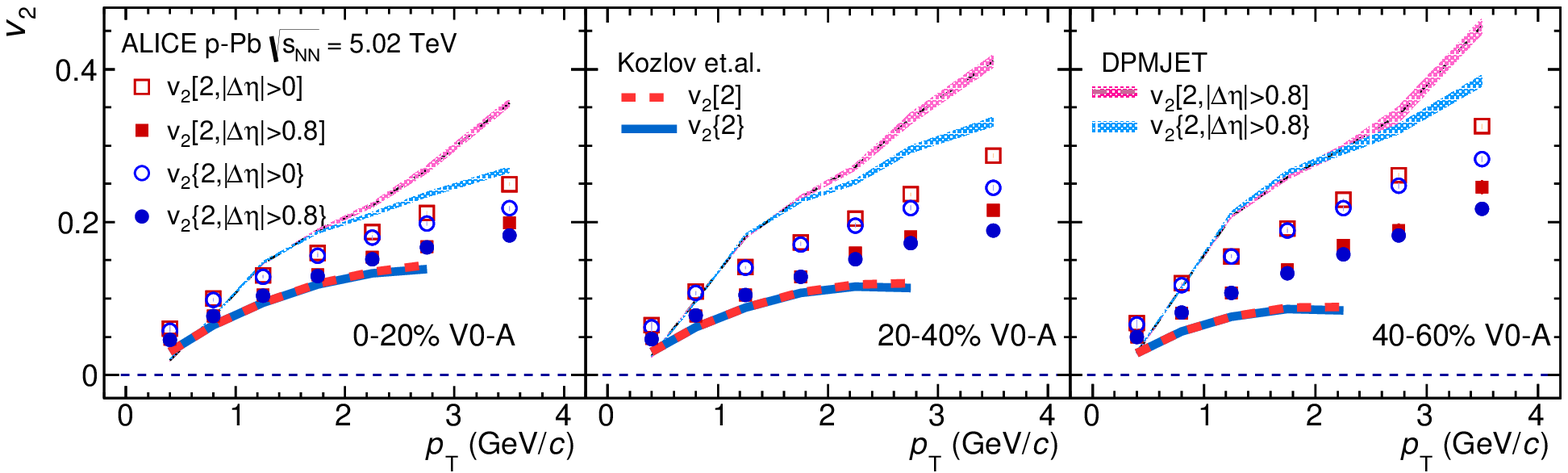}
\caption{$v_{2}\{2, |\Delta\eta| > 0\}$, $v_{2}[2, |\Delta\eta| > 0]$, $v_{2}\{2, |\Delta\eta| > 0.8\}$ and $v_{2}[2, |\Delta\eta| > 0.8]$ for various multiplicity classes of p--Pb collisions at $\sqrt{s_{_{\rm NN}}} = 5.02$~TeV. DPMJET calculations are presented by red shaded lines for $v_{2}\{2, |\Delta\eta| > 0.8\}$ and blue shaded lines for $v_{2}[2, |\Delta\eta| > 0.8]$.
Hydrodynamic calculations (MUSIC)~\cite{Kozlov:2014fqa} with modified MC-Glauber initial conditions and $\eta/s =$ 0.08 for $v_{2}\{2\}$ and $v_{2}[2]$ are shown in solid blue and dashed red lines. }
\label{fig:v22pPbhydro} 
\end{center}
\end{figure}

\begin{figure}[h]
\begin{center}
\includegraphics[width=0.9\textwidth]{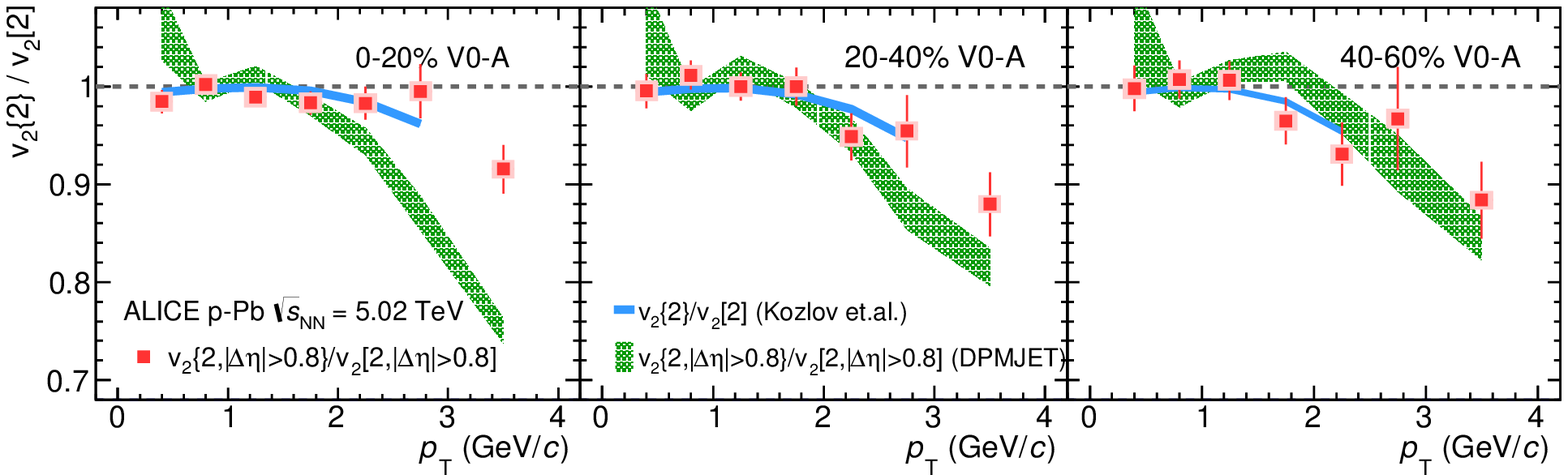}
\caption{The ratio $v_{2}\{2, |\Delta\eta| > 0.8\} / v_{2}[2, |\Delta\eta| > 0.8]$ for various multiplicity classes in p--Pb collisions at $\sqrt{s_{_{\rm NN}}} = 5.02$~TeV. DPMJET calculations are presented by green shaded lines.
Hydrodynamic calculations (MUSIC)~\cite{Kozlov:2014fqa} with modified MC-Glauber initial conditions and $\eta/s =$ 0.08 are shown as solid blue lines.}
\label{fig:v22pPb} 
\end{center}
\end{figure}

Figure~\ref{fig:v22pPbhydro} presents $v_{2}\{2\}$ and $v_{2}[2]$ with $|\Delta\eta|>0$ and $|\Delta\eta|>0.8$ for various multiplicity classes in p--Pb collisions at $\sqrt{s_{_{\rm NN}}} = 5.02$~TeV. 
It is shown that, after applying the pseudorapidity gap $|\Delta\eta|>$ 0.8, both $v_{2}\{2\}$ and $v_{2}[2]$ decrease substantially, in particular for more peripheral collisions, mainly due to the reduction of non-flow effects. 
The ratio $v_{2}\{2, |\Delta\eta| > 0.8\}/v_{2}[2, |\Delta\eta| > 0.8]$, shown in Fig.~\ref{fig:v22pPb}, displays hints of deviations from unity above $\pt \approx 2$ GeV/$c$, but the statistical uncertainties are still too large to draw a firm conclusion. 
The DPMJET model~\cite{Roesler:2000he}, which is an implementation of the two-component Dual Parton Model for the description of interactions involving nuclei, and contains no collective effects, has been used as a benchmark to study the influence of non-flow in p--Pb collisions ~\cite{Abelev:2014mda}. The calculations based on DPMJET simulations are compared to data. It is observed in Fig.~\ref{fig:v22pPbhydro} that DPMJET overestimates $v_{2}$ significantly for the presented multiplicity classes, and generates higher $v_{2}$ coefficients in lower multiplicity regions. Meanwhile, Fig.~\ref{fig:v22pPb} shows that for $v_{2}\{2\} / v_{2}[2]$ the agreement between data and DPMJET is better in low multiplicity p--Pb collisions, where no evidence of anisotropic collectivity is achieved from previous measurements~\cite{Abelev:2014mda,ABELEV:2013wsa}. 
In addition, the hydrodynamic calculations~\cite{Kozlov:2014fqa} from MUSIC v2.0 using a modified MC-Glauber initial state and $\eta/s =$ 0.08 are also presented in Figs.~\ref{fig:v22pPbhydro} and~\ref{fig:v22pPb}. 
These calculations in general underpredict the measured $v_{2}$ coefficients but agree better with the data in high multiplicity than in low multiplicity classes.
It should be emphasized that in contrast to hydrodynamic calculations, the measured $v_{2}\{2\}$ and $v_{2}[2]$ increase (albeit very slightly in particular when the $|\Delta \eta|$ gap is applied) from 0--20\% to 40--60\% multiplicity classes, which indicates that non-flow effects might play a more important role in low multiplicity events. This could explain the increasing deviation between data and hydrodynamic calculations with $\pt$ and towards lower multiplicity classes, shown in Fig.~\ref{fig:v22pPbhydro}. 
The hydrodynamic calculations reproduce the $v_{2}\{2\} / v_{2}[2]$ measurements in the 0--20 \% multiplicity class, which seems to indicate that hydrodynamic collectivity is present in high multiplicity p--Pb collisions. 
However, it is still unclear at the moment why the measured ratio is still reproduced by hydrodynamic calculations for multiplicity class above 20\%, where no significant flow signal is expected to be produced~\cite{Abelev:2014mda}. The agreement might be accidental since the DPMJET and hydrodynamic calculations also agree with each other in this class.

\begin{figure}[thb]
\begin{center}
\includegraphics[width=0.9\textwidth]{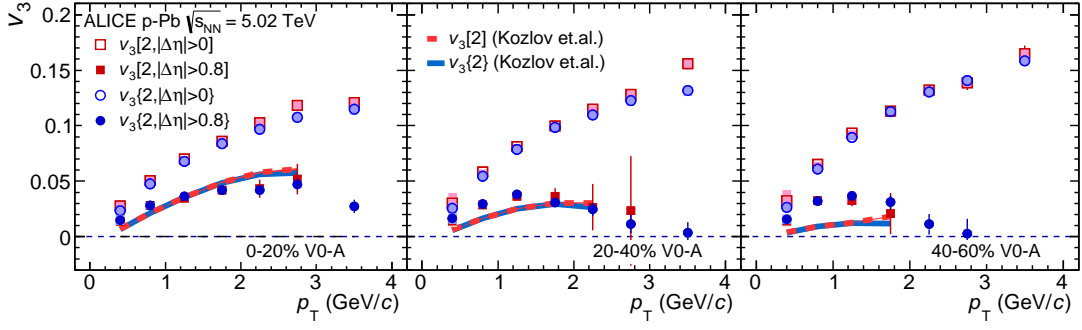}
\caption{$v_{3}\{2, |\Delta\eta| > 0\}$, $v_{3}[2, |\Delta\eta| > 0]$, $v_{3}\{2, |\Delta\eta| > 0.8\}$ and $v_{3}[2, |\Delta\eta| > 0.8]$ for various multiplicity classes in p--Pb collisions at $\sqrt{s_{_{\rm NN}}} = 5.02$~TeV. Hydrodynamic calculations (MUSIC)~\cite{Kozlov:2014fqa} with modified MC-Glauber initial conditions and $\eta/s =$ 0.08 for $v_{2}\{2\}$ and $v_{2}[2]$ are shown as solid blue and dashed red lines.}
\label{fig:v32pPb} 
\end{center}
\end{figure}

The $v_{3}\{2\}$ and $v_{3}[2]$ measured with $|\Delta\eta|>0$ and $|\Delta\eta|>0.8$ in p--Pb collisions at $\sqrt{s_{_{\rm NN}}} = 5.02$ TeV are shown in Fig.~\ref{fig:v32pPb}. 
Both $v_{3}\{2, |\Delta\eta| > 0\}$ and $v_{3}[2, |\Delta\eta| > 0]$ increase with $\pt$ and also with decreasing multiplicity. The measured $v_{3}\{2\}$ and $v_{3}[2]$ with a pseudorapidity gap of $|\Delta\eta|>0.8$ are much smaller than those with $|\Delta\eta|>0$, with the deviation increasing as a function of $\pt$. The relative influence of non-flow effects appears to be stronger in $v_3$ than in $v_2$ measurements. A similar qualitative behaviour was observed for $p_{\rm T}$-integrated two-particle cumulants $c_{2}\{2\}$ and $c_{3}\{2\}$ in p--Pb collisions, measured as functions of multiplicity for different $|\Delta \eta|$ gaps~\cite{ABELEV:2013wsa}. It might be worth noting that part of the remaining non-flow contamination with $|\Delta \eta|>0.8$, the recoil jet ridge, has a positive sign contribution for $v_2$ and a negative sign one for $v_3$ for $\pt > 2$ GeV/$c$.
In addition, it is found that hydrodynamic calculations describe the data better at high multiplicity than at low multiplicity, while DPMJET generates negative ($v_{3}[2])^{2}$ values for all multiplicity classes and thus cannot be shown here for comparison.
Furthermore, the deviations between $v_{3}\{2, |\Delta\eta| > 0.8\}$ and $v_{3}[2, |\Delta\eta| > 0.8]$ are not observed for the presented $p_{\rm T}$ region. There is no indication of $p_{\rm T}$-dependent $V_{3}$ flow vector fluctuations in p--Pb collisions. 

\begin{figure}
\begin{center}
\includegraphics[width=0.75\textwidth]{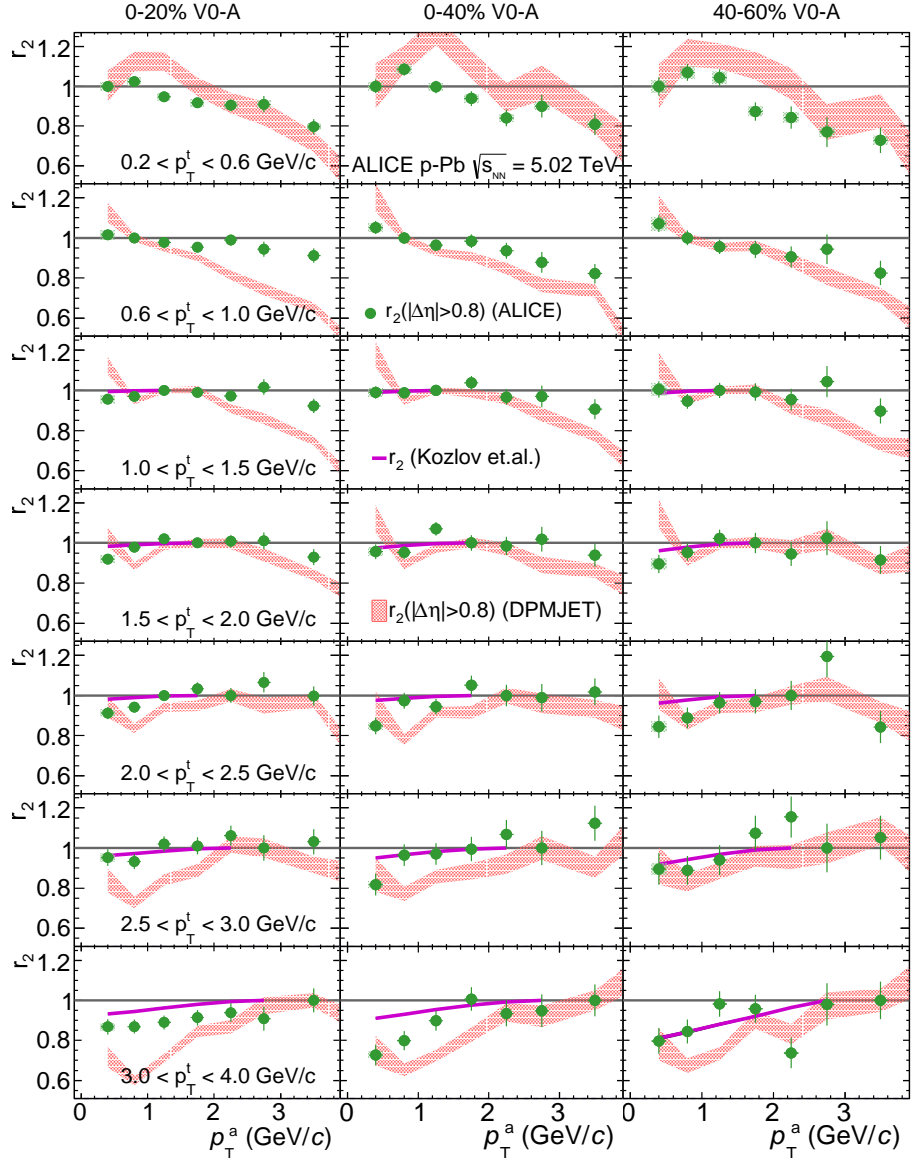}
\caption{ The factorisation ratio $r_{2}$, as a function of $p_{\rm T}^{\, a}$ in bins of $p_{\rm T}^{t}$ for multiplicity classes 0--20$\%$, 20--40$\%$ and 40--60$\%$ in p--Pb collisions at $\sqrt{s_{_{\rm NN}}} = 5.02$~TeV, are presented by solid magenta circles. 
DPMJET calculations are presented by pink shaded areas.
Hydrodynamic calculations (MUSIC) with modified MC-Glauber initial conditions and $\eta/s =$ 0.08 are shown as magenta lines~\cite{Kozlov:2014fqa}.} 
\label{fig:r2pPb} 
\end{center} 
\end{figure}

Figure~\ref{fig:r2pPb} shows $r_{2}(|\Delta\eta|>0.8)$ measurements as a function of $p_{\rm T}^{\, a}$ in three $p_{\rm T}^{t}$ intervals for multiplicity classes 0--20$\%$, 20--40$\%$ and 40--60$\%$ in p--Pb collisions at $\sqrt{s_{_{\rm NN}}} = 5.02$ TeV. 
The $r_{2}(|\Delta\eta|>0.8)$ deviates from unity when the $\pt^{\, t}$ and $\pt^{\, a}$ are well away from each other (most pronouncedly in the lowest and highest $\pt^{\, t}$ bins) with the trend being similar for all multiplicity classes. As mentioned earlier, the deviation is more significant at high multiplicity.
In overlapping $\pt^{\, t}$ and $\pt^{\, a}$ intervals, the $r_{2}$ measurements in the highest multiplicity p--Pb events are consistent with those made by CMS Collaboration~\cite{Khachatryan:2015oea}. The breakdown of factorisation is more pronounced in high multiplicity p--Pb collisions than in the 40--50\% centrality class in Pb--Pb collisions (see Fig.~\ref{fig:r2etagaps}). The DPMJET calculations are presented for comparison. It is clearly seen that DPMJET overestimates the deviations of $r_2$ from unity in the high multiplicity region, nevertheless, the calculation describes the data better in low multiplicity events in which non-flow effects are dominant.
At the same time, these measurements are found to be compatible with hydrodynamic calculations using modified MC-Glauber initial conditions and $\eta/s =$ 0.08.
When selecting the triggered particles from 0.6 $< \pt^{\, t} <$1.0 GeV/$c$ or 1.0 $< \pt^{\, t} <$ 1.5 GeV/$c$, the trend of $r_{2}$ looks similar to that of $v_{2}\{2\} / v_{2}[2]$, mainly because the mean $\pt$ of charged particles is within 0.6 $< \langle \pt \rangle <$ 1.0 GeV/$c$~\cite{Abelev:2013bla}.

\section{Summary}
\label{sec:summary}

Searches for $p_{\rm T}$-dependent flow vector fluctuations are performed by measuring $v_{n}\{2\} / v_{n}[2]$ and $r_{n}$ in Pb--Pb collisions at $\sqrt{s_{_{\rm NN}}}=$ 2.76 TeV and p--Pb collisions at $\sqrt{s_{_{\rm NN}}}=$ 5.02 TeV. 
In Pb--Pb collisions, both $v_{2}\{2\} / v_{2}[2]$ and $r_{2}$ show deviations from unity, and the $r_2$ results are consistent with previous measurements from the CMS Collaboration. These effects are more pronounced in the most central collisions and cannot be explained solely by non-flow effects.
Therefore, these results suggest the presence of possible $V_{2}$ vector fluctuations in Pb--Pb collisions. It further implies that the traditional $v_{2}\{2\}$ results should be interpreted precisely as the correlations of the azimuthal angle of produced particles with respect to the $\pt$ integrated flow vector over a certain kinematic region. Future comparisons between theoretical calculations and experimental measurements should be based on the same kinematic conditions. These comparisons, performed under carefully defined precisely matching kinematic conditions, are crucial to constrain the initial conditions and precisely extract the transport properties of the produced matter, without possible bias from additional $p_{\rm T}$-dependent flow vector fluctuations.
Meanwhile, no significant deviation of $v_{3}\{2\}/v_{3}[2]$ or $v_{4}\{2\}/v_{4}[2]$ from unity was observed, meaning that there is no indication of $\pt$-dependent $V_{3}$ and $V_{4}$ vector fluctuations.
The comparison to hydrodynamic calculations shows only the calculations from $\tt iEBE$-$\tt VISHNU$ with AMPT initial conditions could describe the data quantitatively. The measurements presented in this paper provide a unique approach to constrain the initial conditions and transport properties, e.g. shear viscosity over entropy density ratio $\eta/s$ of the QGP, complementing the previous anisotropic flow measurements. The results therefore bring new insights into the properties of the QGP produced in relativistic heavy ion collisions at the CERN Large Hadron Collider.

Similar studies were performed in various multiplicity classes in p--Pb collisions. Deviations of $v_{2}\{2\} / v_{2}[2]$ and $r_{2}$ from unity are observed, although with relatively large statistical fluctuations. For the highest p--Pb multiplicity class, the deviations are significantly overestimated by DPMJET; however, they are compatible with hydrodynamic calculations using modified MC-Glauber initial conditions and $\eta/s =$ 0.08. Meanwhile for low multiplicity p--Pb collisions, the data sits between calculations from DPMJET and hydrodynamics. Neither the DPMJET model, which does not incorporate anisotropic flow, nor the hydrodynamic model, which does not include non-flow contributions, could provide a quantitative description of the data. Future theoretical developments together with comparisons to high-precision measurements are crucial to give a certain answer concerning $p_{\rm T}$-dependent vector $V_{n}$ fluctuations in p--Pb collisions.

\newenvironment{acknowledgement}{\relax}{\relax}
\begin{acknowledgement}
\section*{Acknowledgements}
\input{fa_2017-05-31.tex}
\end{acknowledgement}

\bibliographystyle{utphys}   
\bibliography{bibliography}

\newpage
\appendix
\section{The ALICE Collaboration}
\label{app:collab}
\input{Alice_Authorlist_2017-May-31.tex}

\end{document}

%% file: fa_2017-05-31.tex

The ALICE Collaboration would like to thank all its engineers and technicians for their invaluable contributions to the construction of the experiment and the CERN accelerator teams for the outstanding performance of the LHC complex.
The ALICE Collaboration gratefully acknowledges the resources and support provided by all Grid centres and the Worldwide LHC Computing Grid (WLCG) collaboration.
The ALICE Collaboration acknowledges the following funding agencies for their support in building and running the ALICE detector:
A. I. Alikhanyan National Science Laboratory (Yerevan Physics Institute) Foundation (ANSL), State Committee of Science and World Federation of Scientists (WFS), Armenia;
Austrian Academy of Sciences and Nationalstiftung f\"{u}r Forschung, Technologie und Entwicklung, Austria;
Ministry of Communications and High Technologies, National Nuclear Research Center, Azerbaijan;
Conselho Nacional de Desenvolvimento Cient\'{\i}fico e Tecnol\'{o}gico (CNPq), Universidade Federal do Rio Grande do Sul (UFRGS), Financiadora de Estudos e Projetos (Finep) and Funda\c{c}\~{a}o de Amparo \`{a} Pesquisa do Estado de S\~{a}o Paulo (FAPESP), Brazil;
Ministry of Science \& Technology of China (MSTC), National Natural Science Foundation of China (NSFC) and Ministry of Education of China (MOEC) , China;
Ministry of Science, Education and Sport and Croatian Science Foundation, Croatia;
Ministry of Education, Youth and Sports of the Czech Republic, Czech Republic;
The Danish Council for Independent Research | Natural Sciences, the Carlsberg Foundation and Danish National Research Foundation (DNRF), Denmark;
Helsinki Institute of Physics (HIP), Finland;
Commissariat \`{a} l'Energie Atomique (CEA) and Institut National de Physique Nucl\'{e}aire et de Physique des Particules (IN2P3) and Centre National de la Recherche Scientifique (CNRS), France;
Bundesministerium f\"{u}r Bildung, Wissenschaft, Forschung und Technologie (BMBF) and GSI Helmholtzzentrum f\"{u}r Schwerionenforschung GmbH, Germany;
General Secretariat for Research and Technology, Ministry of Education, Research and Religions, Greece;
National Research, Development and Innovation Office, Hungary;
Department of Atomic Energy Government of India (DAE) and Council of Scientific and Industrial Research (CSIR), New Delhi, India;
Indonesian Institute of Science, Indonesia;
Centro Fermi - Museo Storico della Fisica e Centro Studi e Ricerche Enrico Fermi and Istituto Nazionale di Fisica Nucleare (INFN), Italy;
Institute for Innovative Science and Technology , Nagasaki Institute of Applied Science (IIST), Japan Society for the Promotion of Science (JSPS) KAKENHI and Japanese Ministry of Education, Culture, Sports, Science and Technology (MEXT), Japan;
Consejo Nacional de Ciencia (CONACYT) y Tecnolog\'{i}a, through Fondo de Cooperaci\'{o}n Internacional en Ciencia y Tecnolog\'{i}a (FONCICYT) and Direcci\'{o}n General de Asuntos del Personal Academico (DGAPA), Mexico;
Nederlandse Organisatie voor Wetenschappelijk Onderzoek (NWO), Netherlands;
The Research Council of Norway, Norway;
Commission on Science and Technology for Sustainable Development in the South (COMSATS), Pakistan;
Pontificia Universidad Cat\'{o}lica del Per\'{u}, Peru;
Ministry of Science and Higher Education and National Science Centre, Poland;
Korea Institute of Science and Technology Information and National Research Foundation of Korea (NRF), Republic of Korea;
Ministry of Education and Scientific Research, Institute of Atomic Physics and Romanian National Agency for Science, Technology and Innovation, Romania;
Joint Institute for Nuclear Research (JINR), Ministry of Education and Science of the Russian Federation and National Research Centre Kurchatov Institute, Russia;
Ministry of Education, Science, Research and Sport of the Slovak Republic, Slovakia;
National Research Foundation of South Africa, South Africa;
Centro de Aplicaciones Tecnol\'{o}gicas y Desarrollo Nuclear (CEADEN), Cubaenerg\'{\i}a, Cuba, Ministerio de Ciencia e Innovacion and Centro de Investigaciones Energ\'{e}ticas, Medioambientales y Tecnol\'{o}gicas (CIEMAT), Spain;
Swedish Research Council (VR) and Knut \& Alice Wallenberg Foundation (KAW), Sweden;
European Organization for Nuclear Research, Switzerland;
National Science and Technology Development Agency (NSDTA), Suranaree University of Technology (SUT) and Office of the Higher Education Commission under NRU project of Thailand, Thailand;
Turkish Atomic Energy Agency (TAEK), Turkey;
National Academy of  Sciences of Ukraine, Ukraine;
Science and Technology Facilities Council (STFC), United Kingdom;
National Science Foundation of the United States of America (NSF) and United States Department of Energy, Office of Nuclear Physics (DOE NP), United States of America.

%% file: Alice_Authorlist_2017-May-31.tex


\begingroup
\small
\begin{flushleft}
S.~Acharya$^\textrm{\scriptsize 139}$,
D.~Adamov\'{a}$^\textrm{\scriptsize 96}$,
J.~Adolfsson$^\textrm{\scriptsize 34}$,
M.M.~Aggarwal$^\textrm{\scriptsize 101}$,
G.~Aglieri Rinella$^\textrm{\scriptsize 35}$,
M.~Agnello$^\textrm{\scriptsize 31}$,
N.~Agrawal$^\textrm{\scriptsize 48}$,
Z.~Ahammed$^\textrm{\scriptsize 139}$,
N.~Ahmad$^\textrm{\scriptsize 17}$,
S.U.~Ahn$^\textrm{\scriptsize 80}$
% ...
\renewcommand\labelenumi{\textsuperscript{\theenumi}~}

\section*{Affiliation notes}
\renewcommand\theenumi{\roman{enumi}}
\begin{Authlist}
\item \Adef{0}Deceased
\item \Adef{idp1789808}{Also at: Dipartimento DET del Politecnico di Torino, Turin, Italy}
\item \Adef{idp1809200}{Also at: Georgia State University, Atlanta, Georgia, United States}
\item \Adef{idp4118176}{Also at: M.V. Lomonosov Moscow State University, D.V. Skobeltsyn Institute of Nuclear, Physics, Moscow, Russia}
\item \Adef{idp4476496}{Also at: Department of Applied Physics, Aligarh Muslim University, Aligarh, India}
\item \Adef{idp5453424}{Also at: Institute of Theoretical Physics, University of Wroclaw, Poland}
\end{Authlist}

\section*{Collaboration Institutes}
\renewcommand\theenumi{\arabic{enumi}~}

$^{1}$A.I. Alikhanyan National Science Laboratory (Yerevan Physics Institute) Foundation, Yerevan, Armenia
\\
$^{2}$Benem\'{e}rita Universidad Aut\'{o}noma de Puebla, Puebla, Mexico
\\
$^{3}$Bogolyubov Institute for Theoretical Physics, Kiev, Ukraine
\\
$^{4}$Bose Institute, Department of Physics 
and Centre for Astroparticle Physics and Space Science (CAPSS), Kolkata, India
\\
$^{5}$Budker Institute for Nuclear Physics, Novosibirsk, Russia
\\
$^{6}$California Polytechnic State University, San Luis Obispo, California, United States
\\
$^{7}$Central China Normal University, Wuhan, China
\\
$^{8}$Centre de Calcul de l'IN2P3, Villeurbanne, Lyon, France
\\
$^{9}$Centro de Aplicaciones Tecnol\'{o}gicas y Desarrollo Nuclear (CEADEN), Havana, Cuba
\\
$^{10}$Centro de Investigaciones Energ\'{e}ticas Medioambientales y Tecnol\'{o}gicas (CIEMAT), Madrid, Spain
\\
$^{11}$Centro de Investigaci\'{o}n y de Estudios Avanzados (CINVESTAV), Mexico City and M\'{e}rida, Mexico
\\
$^{12}$Centro Fermi - Museo Storico della Fisica e Centro Studi e Ricerche ``Enrico Fermi', Rome, Italy
\\
$^{13}$Chicago State University, Chicago, Illinois, United States
\\
$^{14}$China Institute of Atomic Energy, Beijing, China
\\
$^{15}$COMSATS Institute of Information Technology (CIIT), Islamabad, Pakistan
\\
$^{16}$Departamento de F\'{\i}sica de Part\'{\i}culas and IGFAE, Universidad de Santiago de Compostela, Santiago de Compostela, Spain
\\
$^{17}$Department of Physics, Aligarh Muslim University, Aligarh, India
\\
$^{18}$Department of Physics, Ohio State University, Columbus, Ohio, United States
\\
$^{19}$Department of Physics, Pusan National University, Pusan, South Korea
\\
$^{20}$Department of Physics, Sejong University, Seoul, South Korea
\\
$^{21}$Department of Physics, University of Oslo, Oslo, Norway
\\
$^{22}$Department of Physics and Technology, University of Bergen, Bergen, Norway
\\
$^{23}$Dipartimento di Fisica dell'Universit\`{a} 'La Sapienza'
and Sezione INFN, Rome, Italy
\\
$^{24}$Dipartimento di Fisica dell'Universit\`{a}
and Sezione INFN, Cagliari, Italy
\\
$^{25}$Dipartimento di Fisica dell'Universit\`{a}
and Sezione INFN, Trieste, Italy
\\
$^{26}$Dipartimento di Fisica dell'Universit\`{a}
and Sezione INFN, Turin, Italy
\\
$^{27}$Dipartimento di Fisica e Astronomia dell'Universit\`{a}
and Sezione INFN, Bologna, Italy
\\
$^{28}$Dipartimento di Fisica e Astronomia dell'Universit\`{a}
and Sezione INFN, Catania, Italy
\\
$^{29}$Dipartimento di Fisica e Astronomia dell'Universit\`{a}
and Sezione INFN, Padova, Italy
\\
$^{30}$Dipartimento di Fisica `E.R.~Caianiello' dell'Universit\`{a}
and Gruppo Collegato INFN, Salerno, Italy
\\
$^{31}$Dipartimento DISAT del Politecnico and Sezione INFN, Turin, Italy
\\
$^{32}$Dipartimento di Scienze e Innovazione Tecnologica dell'Universit\`{a} del Piemonte Orientale and INFN Sezione di Torino, Alessandria, Italy
\\
$^{33}$Dipartimento Interateneo di Fisica `M.~Merlin'
and Sezione INFN, Bari, Italy
\\
$^{34}$Division of Experimental High Energy Physics, University of Lund, Lund, Sweden
\\
$^{35}$European Organization for Nuclear Research (CERN), Geneva, Switzerland
\\
$^{36}$Excellence Cluster Universe, Technische Universit\"{a}t M\"{u}nchen, Munich, Germany
\\
$^{37}$Faculty of Engineering, Bergen University College, Bergen, Norway
\\
$^{38}$Faculty of Mathematics, Physics and Informatics, Comenius University, Bratislava, Slovakia
\\
$^{39}$Faculty of Nuclear Sciences and Physical Engineering, Czech Technical University in Prague, Prague, Czech Republic
\\
$^{40}$Faculty of Science, P.J.~\v{S}af\'{a}rik University, Ko\v{s}ice, Slovakia
\\
$^{41}$Faculty of Technology, Buskerud and Vestfold University College, Tonsberg, Norway
\\
$^{42}$Frankfurt Institute for Advanced Studies, Johann Wolfgang Goethe-Universit\"{a}t Frankfurt, Frankfurt, Germany
\\
$^{43}$Gangneung-Wonju National University, Gangneung, South Korea
\\
$^{44}$Gauhati University, Department of Physics, Guwahati, India
\\
$^{45}$Helmholtz-Institut f\"{u}r Strahlen- und Kernphysik, Rheinische Friedrich-Wilhelms-Universit\"{a}t Bonn, Bonn, Germany
\\
$^{46}$Helsinki Institute of Physics (HIP), Helsinki, Finland
\\
$^{47}$Hiroshima University, Hiroshima, Japan
\\
$^{48}$Indian Institute of Technology Bombay (IIT), Mumbai, India
\\
$^{49}$Indian Institute of Technology Indore, Indore, India
\\
$^{50}$Indonesian Institute of Sciences, Jakarta, Indonesia
\\
$^{51}$INFN, Laboratori Nazionali di Frascati, Frascati, Italy
\\
$^{52}$INFN, Laboratori Nazionali di Legnaro, Legnaro, Italy
\\
$^{53}$INFN, Sezione di Bari, Bari, Italy
\\
$^{54}$INFN, Sezione di Bologna, Bologna, Italy
\\
$^{55}$INFN, Sezione di Cagliari, Cagliari, Italy
\\
$^{56}$INFN, Sezione di Catania, Catania, Italy
\\
$^{57}$INFN, Sezione di Padova, Padova, Italy
\\
$^{58}$INFN, Sezione di Roma, Rome, Italy
\\
$^{59}$INFN, Sezione di Torino, Turin, Italy
\\
$^{60}$INFN, Sezione di Trieste, Trieste, Italy
\\
$^{61}$Inha University, Incheon, South Korea
\\
$^{62}$Institut de Physique Nucl\'eaire d'Orsay (IPNO), Universit\'e Paris-Sud, CNRS-IN2P3, Orsay, France
\\
$^{63}$Institute for Nuclear Research, Academy of Sciences, Moscow, Russia
\\
$^{64}$Institute for Subatomic Physics of Utrecht University, Utrecht, Netherlands
\\
$^{65}$Institute for Theoretical and Experimental Physics, Moscow, Russia
\\
$^{66}$Institute of Experimental Physics, Slovak Academy of Sciences, Ko\v{s}ice, Slovakia
\\
$^{67}$Institute of Physics, Academy of Sciences of the Czech Republic, Prague, Czech Republic
\\
$^{68}$Institute of Physics, Bhubaneswar, India
\\
$^{69}$Institute of Space Science (ISS), Bucharest, Romania
\\
$^{70}$Institut f\"{u}r Informatik, Johann Wolfgang Goethe-Universit\"{a}t Frankfurt, Frankfurt, Germany
\\
$^{71}$Institut f\"{u}r Kernphysik, Johann Wolfgang Goethe-Universit\"{a}t Frankfurt, Frankfurt, Germany
\\
$^{72}$Institut f\"{u}r Kernphysik, Westf\"{a}lische Wilhelms-Universit\"{a}t M\"{u}nster, M\"{u}nster, Germany
\\
$^{73}$Instituto de Ciencias Nucleares, Universidad Nacional Aut\'{o}noma de M\'{e}xico, Mexico City, Mexico
\\
$^{74}$Instituto de F\'{i}sica, Universidade Federal do Rio Grande do Sul (UFRGS), Porto Alegre, Brazil
\\
$^{75}$Instituto de F\'{\i}sica, Universidad Nacional Aut\'{o}noma de M\'{e}xico, Mexico City, Mexico
\\
$^{76}$IRFU, CEA, Universit\'{e} Paris-Saclay, Saclay, France
\\
$^{77}$iThemba LABS, National Research Foundation, Somerset West, South Africa
\\
$^{78}$Joint Institute for Nuclear Research (JINR), Dubna, Russia
\\
$^{79}$Konkuk University, Seoul, South Korea
\\
$^{80}$Korea Institute of Science and Technology Information, Daejeon, South Korea
\\
$^{81}$KTO Karatay University, Konya, Turkey
\\
$^{82}$Laboratoire de Physique Corpusculaire (LPC), Clermont Universit\'{e}, Universit\'{e} Blaise Pascal, CNRS--IN2P3, Clermont-Ferrand, France
\\
$^{83}$Laboratoire de Physique Subatomique et de Cosmologie, Universit\'{e} Grenoble-Alpes, CNRS-IN2P3, Grenoble, France
\\
$^{84}$Lawrence Berkeley National Laboratory, Berkeley, California, United States
\\
$^{85}$Moscow Engineering Physics Institute, Moscow, Russia
\\
$^{86}$Nagasaki Institute of Applied Science, Nagasaki, Japan
\\
$^{87}$National and Kapodistrian University of Athens, Physics Department, Athens, Greece
\\
$^{88}$National Centre for Nuclear Studies, Warsaw, Poland
\\
$^{89}$National Institute for Physics and Nuclear Engineering, Bucharest, Romania
\\
$^{90}$National Institute of Science Education and Research, Bhubaneswar, India
\\
$^{91}$National Nuclear Research Center, Baku, Azerbaijan
\\
$^{92}$National Research Centre Kurchatov Institute, Moscow, Russia
\\
$^{93}$Niels Bohr Institute, University of Copenhagen, Copenhagen, Denmark
\\
$^{94}$Nikhef, Nationaal instituut voor subatomaire fysica, Amsterdam, Netherlands
\\
$^{95}$Nuclear Physics Group, STFC Daresbury Laboratory, Daresbury, United Kingdom
\\
$^{96}$Nuclear Physics Institute, Academy of Sciences of the Czech Republic, \v{R}e\v{z} u Prahy, Czech Republic
\\
$^{97}$Oak Ridge National Laboratory, Oak Ridge, Tennessee, United States
\\
$^{98}$Petersburg Nuclear Physics Institute, Gatchina, Russia
\\
$^{99}$Physics Department, Creighton University, Omaha, Nebraska, United States
\\
$^{100}$Physics department, Faculty of science, University of Zagreb, Zagreb, Croatia
\\
$^{101}$Physics Department, Panjab University, Chandigarh, India
\\
$^{102}$Physics Department, University of Cape Town, Cape Town, South Africa
\\
$^{103}$Physics Department, University of Jammu, Jammu, India
\\
$^{104}$Physics Department, University of Rajasthan, Jaipur, India
\\
$^{105}$Physikalisches Institut, Eberhard Karls Universit\"{a}t T\"{u}bingen, T\"{u}bingen, Germany
\\
$^{106}$Physikalisches Institut, Ruprecht-Karls-Universit\"{a}t Heidelberg, Heidelberg, Germany
\\
$^{107}$Physik Department, Technische Universit\"{a}t M\"{u}nchen, Munich, Germany
\\
$^{108}$Purdue University, West Lafayette, Indiana, United States
\\
$^{109}$Research Division and ExtreMe Matter Institute EMMI, GSI Helmholtzzentrum f\"ur Schwerionenforschung GmbH, Darmstadt, Germany
\\
$^{110}$Rudjer Bo\v{s}kovi\'{c} Institute, Zagreb, Croatia
\\
$^{111}$Russian Federal Nuclear Center (VNIIEF), Sarov, Russia
\\
$^{112}$Saha Institute of Nuclear Physics, Kolkata, India
\\
$^{113}$School of Physics and Astronomy, University of Birmingham, Birmingham, United Kingdom
\\
$^{114}$Secci\'{o}n F\'{\i}sica, Departamento de Ciencias, Pontificia Universidad Cat\'{o}lica del Per\'{u}, Lima, Peru
\\
$^{115}$SSC IHEP of NRC Kurchatov institute, Protvino, Russia
\\
$^{116}$Stefan Meyer Institut f\"{u}r Subatomare Physik (SMI), Vienna, Austria
\\
$^{117}$SUBATECH, IMT Atlantique, Universit\'{e} de Nantes, CNRS-IN2P3, Nantes, France
\\
$^{118}$Suranaree University of Technology, Nakhon Ratchasima, Thailand
\\
$^{119}$Technical University of Ko\v{s}ice, Ko\v{s}ice, Slovakia
\\
$^{120}$Technical University of Split FESB, Split, Croatia
\\
$^{121}$The Henryk Niewodniczanski Institute of Nuclear Physics, Polish Academy of Sciences, Cracow, Poland
\\
$^{122}$The University of Texas at Austin, Physics Department, Austin, Texas, United States
\\
$^{123}$Universidad Aut\'{o}noma de Sinaloa, Culiac\'{a}n, Mexico
\\
$^{124}$Universidade de S\~{a}o Paulo (USP), S\~{a}o Paulo, Brazil
\\
$^{125}$Universidade Estadual de Campinas (UNICAMP), Campinas, Brazil
\\
$^{126}$Universidade Federal do ABC, Santo Andre, Brazil
\\
$^{127}$University of Houston, Houston, Texas, United States
\\
$^{128}$University of Jyv\"{a}skyl\"{a}, Jyv\"{a}skyl\"{a}, Finland
\\
$^{129}$University of Liverpool, Liverpool, United Kingdom
\\
$^{130}$University of Tennessee, Knoxville, Tennessee, United States
\\
$^{131}$University of the Witwatersrand, Johannesburg, South Africa
\\
$^{132}$University of Tokyo, Tokyo, Japan
\\
$^{133}$University of Tsukuba, Tsukuba, Japan
\\
$^{134}$Universit\'{e} de Lyon, Universit\'{e} Lyon 1, CNRS/IN2P3, IPN-Lyon, Villeurbanne, Lyon, France
\\
$^{135}$Universit\'{e} de Strasbourg, CNRS, IPHC UMR 7178, F-67000 Strasbourg, France, Strasbourg, France
\\
$^{136}$Universit\`{a} degli Studi di Pavia, Pavia, Italy
\\
$^{137}$Universit\`{a} di Brescia, Brescia, Italy
\\
$^{138}$V.~Fock Institute for Physics, St. Petersburg State University, St. Petersburg, Russia
\\
$^{139}$Variable Energy Cyclotron Centre, Kolkata, India
\\
$^{140}$Warsaw University of Technology, Warsaw, Poland
\\
$^{141}$Wayne State University, Detroit, Michigan, United States
\\
$^{142}$Wigner Research Centre for Physics, Hungarian Academy of Sciences, Budapest, Hungary
\\
$^{143}$Yale University, New Haven, Connecticut, United States
\\
$^{144}$Yonsei University, Seoul, South Korea
\\
$^{145}$Zentrum f\"{u}r Technologietransfer und Telekommunikation (ZTT), Fachhochschule Worms, Worms, Germany
\endgroup

%% file: alicepreprint_CDS_20170731.bbl
\providecommand{\href}[2]{#2}\begingroup\raggedright\begin{thebibliography}{10}

\bibitem{Shuryak:1978ij}
E.~V. Shuryak, ``{Quark-Gluon Plasma and Hadronic Production of Leptons,
  Photons and Psions},''
\href{http://dx.doi.org/10.1016/0370-2693(78)90370-2}{{\em Phys. Lett.}
  {\bfseries B78} (1978) 150}.

\bibitem{Shuryak:1980tp}
E.~V. Shuryak, ``{Quantum Chromodynamics and the Theory of Superdense
  Matter},''
\href{http://dx.doi.org/10.1016/0370-1573(80)90105-2}{{\em Phys. Rept.}
  {\bfseries 61} (1980) 71--158}.

\bibitem{Ollitrault:1992bk}
J.-Y. Ollitrault, ``{Anisotropy as a signature of transverse collective
  flow},''
\href{http://dx.doi.org/10.1103/PhysRevD.46.229}{{\em Phys. Rev.} {\bfseries
  D46} (1992) 229--245}.

\bibitem{Voloshin:2008dg}
S.~A. Voloshin, A.~M. Poskanzer, and R.~Snellings, ``{Collective phenomena in
  non-central nuclear collisions},''
\href{http://arxiv.org/abs/0809.2949}{{\ttfamily arXiv:0809.2949 [nucl-ex]}}.

\bibitem{Heinz:2013th}
U.~Heinz and R.~Snellings, ``{Collective flow and viscosity in relativistic
  heavy-ion collisions},''
  \href{http://dx.doi.org/10.1146/annurev-nucl-102212-170540}{{\em Ann. Rev.
  Nucl. Part. Sci.} {\bfseries 63} (2013) 123--151},
\href{http://arxiv.org/abs/1301.2826}{{\ttfamily arXiv:1301.2826 [nucl-th]}}.

\bibitem{Voloshin:1994mz}
S.~Voloshin and Y.~Zhang, ``{Flow study in relativistic nuclear collisions by
  Fourier expansion of Azimuthal particle distributions},''
  \href{http://dx.doi.org/10.1007/s002880050141}{{\em Z. Phys.} {\bfseries C70}
  (1996) 665--672},
\href{http://arxiv.org/abs/hep-ph/9407282}{{\ttfamily arXiv:hep-ph/9407282
  [hep-ph]}}.

\bibitem{Poskanzer:1998yz}
A.~M. Poskanzer and S.~A. Voloshin, ``{Methods for analyzing anisotropic flow
  in relativistic nuclear collisions},''
  \href{http://dx.doi.org/10.1103/PhysRevC.58.1671}{{\em Phys. Rev.} {\bfseries
  C58} (1998) 1671--1678},
\href{http://arxiv.org/abs/nucl-ex/9805001}{{\ttfamily arXiv:nucl-ex/9805001
  [nucl-ex]}}.

\bibitem{Mishra:2007tw}
A.~P. Mishra, R.~K. Mohapatra, P.~S. Saumia, and A.~M. Srivastava,
  ``{Super-horizon fluctuations and acoustic oscillations in relativistic
  heavy-ion collisions},''
  \href{http://dx.doi.org/10.1103/PhysRevC.77.064902}{{\em Phys. Rev.}
  {\bfseries C77} (2008) 064902},
\href{http://arxiv.org/abs/0711.1323}{{\ttfamily arXiv:0711.1323 [hep-ph]}}.

\bibitem{Mishra:2008dm}
A.~P. Mishra, R.~K. Mohapatra, P.~S. Saumia, and A.~M. Srivastava, ``{Using
  cosmic microwave background radiation analysis tools for flow anisotropies in
  relativistic heavy-ion collisions},''
  \href{http://dx.doi.org/10.1103/PhysRevC.81.034903}{{\em Phys. Rev.}
  {\bfseries C81} (2010) 034903},
\href{http://arxiv.org/abs/0811.0292}{{\ttfamily arXiv:0811.0292 [hep-ph]}}.

\bibitem{Takahashi:2009na}
J.~Takahashi, B.~M. Tavares, W.~L. Qian, R.~Andrade, F.~Grassi, Y.~Hama,
  T.~Kodama, and N.~Xu, ``{Topology studies of hydrodynamics using two particle
  correlation analysis},''
  \href{http://dx.doi.org/10.1103/PhysRevLett.103.242301}{{\em Phys. Rev.
  Lett.} {\bfseries 103} (2009) 242301},
\href{http://arxiv.org/abs/0902.4870}{{\ttfamily arXiv:0902.4870 [nucl-th]}}.

\bibitem{Alver:2010gr}
B.~Alver and G.~Roland, ``{Collision geometry fluctuations and triangular flow
  in heavy-ion collisions},''
  \href{http://dx.doi.org/10.1103/PhysRevC.82.039903,
  10.1103/PhysRevC.81.054905}{{\em Phys. Rev.} {\bfseries C81} (2010) 054905},
  \href{http://arxiv.org/abs/1003.0194}{{\ttfamily arXiv:1003.0194 [nucl-th]}}.
[Erratum: Phys. Rev.C82,039903(2010)].

\bibitem{Alver:2010dn}
B.~H. Alver, C.~Gombeaud, M.~Luzum, and J.-Y. Ollitrault, ``{Triangular flow in
  hydrodynamics and transport theory},''
  \href{http://dx.doi.org/10.1103/PhysRevC.82.034913}{{\em Phys. Rev.}
  {\bfseries C82} (2010) 034913},
\href{http://arxiv.org/abs/1007.5469}{{\ttfamily arXiv:1007.5469 [nucl-th]}}.

\bibitem{Teaney:2010vd}
D.~Teaney and L.~Yan, ``{Triangularity and Dipole Asymmetry in Heavy Ion
  Collisions},'' \href{http://dx.doi.org/10.1103/PhysRevC.83.064904}{{\em Phys.
  Rev.} {\bfseries C83} (2011) 064904},
\href{http://arxiv.org/abs/1010.1876}{{\ttfamily arXiv:1010.1876 [nucl-th]}}.

\bibitem{Luzum:2010sp}
M.~Luzum, ``{Collective flow and long-range correlations in relativistic heavy
  ion collisions},''
  \href{http://dx.doi.org/10.1016/j.physletb.2011.01.013}{{\em Phys. Lett.}
  {\bfseries B696} (2011) 499--504},
\href{http://arxiv.org/abs/1011.5773}{{\ttfamily arXiv:1011.5773 [nucl-th]}}.

\bibitem{Arsene:2004fa}
{\bfseries BRAHMS} Collaboration, I.~Arsene {\em et~al.}, ``{Quark gluon plasma
  and color glass condensate at RHIC? The Perspective from the BRAHMS
  experiment},'' \href{http://dx.doi.org/10.1016/j.nuclphysa.2005.02.130}{{\em
  Nucl. Phys.} {\bfseries A757} (2005) 1--27},
\href{http://arxiv.org/abs/nucl-ex/0410020}{{\ttfamily arXiv:nucl-ex/0410020
  [nucl-ex]}}.

\bibitem{Back:2004je}
{\bfseries PHOBOS} Collaboration, B.~B. Back {\em et~al.}, ``{The PHOBOS
  perspective on discoveries at RHIC},''
  \href{http://dx.doi.org/10.1016/j.nuclphysa.2005.03.084}{{\em Nucl. Phys.}
  {\bfseries A757} (2005) 28--101},
\href{http://arxiv.org/abs/nucl-ex/0410022}{{\ttfamily arXiv:nucl-ex/0410022
  [nucl-ex]}}.

\bibitem{Adams:2005dq}
{\bfseries STAR} Collaboration, J.~Adams {\em et~al.}, ``{Experimental and
  theoretical challenges in the search for the quark gluon plasma: The STAR
  Collaboration's critical assessment of the evidence from RHIC collisions},''
  \href{http://dx.doi.org/10.1016/j.nuclphysa.2005.03.085}{{\em Nucl. Phys.}
  {\bfseries A757} (2005) 102--183},
\href{http://arxiv.org/abs/nucl-ex/0501009}{{\ttfamily arXiv:nucl-ex/0501009
  [nucl-ex]}}.

\bibitem{Adcox:2004mh}
{\bfseries PHENIX} Collaboration, K.~Adcox {\em et~al.}, ``{Formation of dense
  partonic matter in relativistic nucleus-nucleus collisions at RHIC:
  Experimental evaluation by the PHENIX collaboration},''
  \href{http://dx.doi.org/10.1016/j.nuclphysa.2005.03.086}{{\em Nucl. Phys.}
  {\bfseries A757} (2005) 184--283},
\href{http://arxiv.org/abs/nucl-ex/0410003}{{\ttfamily arXiv:nucl-ex/0410003
  [nucl-ex]}}.

\bibitem{Aamodt:2010pa}
{\bfseries ALICE} Collaboration, K.~Aamodt {\em et~al.}, ``{Elliptic flow of
  charged particles in Pb-Pb collisions at 2.76 TeV},''
  \href{http://dx.doi.org/10.1103/PhysRevLett.105.252302}{{\em Phys. Rev.
  Lett.} {\bfseries 105} (2010) 252302},
\href{http://arxiv.org/abs/1011.3914}{{\ttfamily arXiv:1011.3914 [nucl-ex]}}.

\bibitem{ALICE:2011ab}
{\bfseries ALICE} Collaboration, K.~Aamodt {\em et~al.}, ``{Higher harmonic
  anisotropic flow measurements of charged particles in Pb-Pb collisions at
  $\sqrt{s_{NN}}$=2.76 TeV},''
  \href{http://dx.doi.org/10.1103/PhysRevLett.107.032301}{{\em Phys. Rev.
  Lett.} {\bfseries 107} (2011) 032301},
\href{http://arxiv.org/abs/1105.3865}{{\ttfamily arXiv:1105.3865 [nucl-ex]}}.

\bibitem{Abelev:2014pua}
{\bfseries ALICE} Collaboration, B.~B. Abelev {\em et~al.}, ``{Elliptic flow of
  identified hadrons in Pb-Pb collisions at $ \sqrt{s_{\mathrm{NN}}}=2.76 $
  TeV},'' \href{http://dx.doi.org/10.1007/JHEP06(2015)190}{{\em JHEP}
  {\bfseries 06} (2015) 190},
\href{http://arxiv.org/abs/1405.4632}{{\ttfamily arXiv:1405.4632 [nucl-ex]}}.

\bibitem{Adam:2016izf}
{\bfseries ALICE} Collaboration, J.~Adam {\em et~al.}, ``{Anisotropic flow of
  charged particles in Pb-Pb collisions at $\sqrt{s_{\rm NN}}=5.02$ TeV},''
  \href{http://dx.doi.org/10.1103/PhysRevLett.116.132302}{{\em Phys. Rev.
  Lett.} {\bfseries 116} no.~13, (2016) 132302},
\href{http://arxiv.org/abs/1602.01119}{{\ttfamily arXiv:1602.01119 [nucl-ex]}}.

\bibitem{Acharya:2017zfg}
{\bfseries ALICE} Collaboration, S.~Acharya {\em et~al.}, ``{Linear and
  non-linear flow modes in Pb-Pb collisions at $\sqrt{s_{\rm NN}} =$ 2.76
  TeV},''
\href{http://arxiv.org/abs/1705.04377}{{\ttfamily arXiv:1705.04377 [nucl-ex]}}.

\bibitem{ATLAS:2011ah}
{\bfseries ATLAS} Collaboration, G.~Aad {\em et~al.}, ``{Measurement of the
  pseudorapidity and transverse momentum dependence of the elliptic flow of
  charged particles in lead-lead collisions at $\sqrt{s_{NN}}=2.76$ TeV with
  the ATLAS detector},''
  \href{http://dx.doi.org/10.1016/j.physletb.2011.12.056}{{\em Phys. Lett.}
  {\bfseries B707} (2012) 330--348},
\href{http://arxiv.org/abs/1108.6018}{{\ttfamily arXiv:1108.6018 [hep-ex]}}.

\bibitem{ATLAS:2012at}
{\bfseries ATLAS} Collaboration, G.~Aad {\em et~al.}, ``{Measurement of the
  azimuthal anisotropy for charged particle production in $\sqrt{s_{NN}}=2.76$
  TeV lead-lead collisions with the ATLAS detector},''
  \href{http://dx.doi.org/10.1103/PhysRevC.86.014907}{{\em Phys. Rev.}
  {\bfseries C86} (2012) 014907},
\href{http://arxiv.org/abs/1203.3087}{{\ttfamily arXiv:1203.3087 [hep-ex]}}.

\bibitem{Aad:2013xma}
{\bfseries ATLAS} Collaboration, G.~Aad {\em et~al.}, ``{Measurement of the
  distributions of event-by-event flow harmonics in lead-lead collisions at =
  2.76 TeV with the ATLAS detector at the LHC},''
  \href{http://dx.doi.org/10.1007/JHEP11(2013)183}{{\em JHEP} {\bfseries 11}
  (2013) 183},
\href{http://arxiv.org/abs/1305.2942}{{\ttfamily arXiv:1305.2942 [hep-ex]}}.

\bibitem{Chatrchyan:2012wg}
{\bfseries CMS} Collaboration, S.~Chatrchyan {\em et~al.}, ``{Centrality
  dependence of dihadron correlations and azimuthal anisotropy harmonics in
  PbPb collisions at $\sqrt{s_{NN}}=2.76$ TeV},''
  \href{http://dx.doi.org/10.1140/epjc/s10052-012-2012-3}{{\em Eur. Phys. J.}
  {\bfseries C72} (2012) 2012},
\href{http://arxiv.org/abs/1201.3158}{{\ttfamily arXiv:1201.3158 [nucl-ex]}}.

\bibitem{Chatrchyan:2012ta}
{\bfseries CMS} Collaboration, S.~Chatrchyan {\em et~al.}, ``{Measurement of
  the elliptic anisotropy of charged particles produced in PbPb collisions at
  $\sqrt{s}_{NN}$=2.76 TeV},''
  \href{http://dx.doi.org/10.1103/PhysRevC.87.014902}{{\em Phys. Rev.}
  {\bfseries C87} no.~1, (2013) 014902},
\href{http://arxiv.org/abs/1204.1409}{{\ttfamily arXiv:1204.1409 [nucl-ex]}}.

\bibitem{Chatrchyan:2012xq}
{\bfseries CMS} Collaboration, S.~Chatrchyan {\em et~al.}, ``{Azimuthal
  anisotropy of charged particles at high transverse momenta in PbPb collisions
  at $\sqrt{s_{NN}}=2.76$ TeV},''
  \href{http://dx.doi.org/10.1103/PhysRevLett.109.022301}{{\em Phys. Rev.
  Lett.} {\bfseries 109} (2012) 022301},
\href{http://arxiv.org/abs/1204.1850}{{\ttfamily arXiv:1204.1850 [nucl-ex]}}.

\bibitem{ALICE:2016kpq}
{\bfseries ALICE} Collaboration, J.~Adam {\em et~al.}, ``{Correlated
  event-by-event fluctuations of flow harmonics in Pb-Pb collisions at
  $\sqrt{s_{_{\rm NN}}}=2.76$ TeV},''
  \href{http://dx.doi.org/10.1103/PhysRevLett.117.182301}{{\em Phys. Rev.
  Lett.} {\bfseries 117} (2016) 182301},
\href{http://arxiv.org/abs/1604.07663}{{\ttfamily arXiv:1604.07663 [nucl-ex]}}.

\bibitem{Song:2017wtw}
H.~Song, Y.~Zhou, and K.~Gajdosova, ``{Collective flow and hydrodynamics in
  large and small systems at the LHC},''
  \href{http://dx.doi.org/10.1007/s41365-017-0245-4}{{\em Nucl. Sci. Tech.}
  {\bfseries 28} no.~7, (2017) 99},
\href{http://arxiv.org/abs/1703.00670}{{\ttfamily arXiv:1703.00670 [nucl-th]}}.

\bibitem{Gardim:2012im}
F.~G. Gardim, F.~Grassi, M.~Luzum, and J.-Y. Ollitrault, ``{Breaking of
  factorization of two-particle correlations in hydrodynamics},''
  \href{http://dx.doi.org/10.1103/PhysRevC.87.031901}{{\em Phys. Rev.}
  {\bfseries C87} no.~3, (2013) 031901},
\href{http://arxiv.org/abs/1211.0989}{{\ttfamily arXiv:1211.0989 [nucl-th]}}.

\bibitem{Heinz:2013bua}
U.~Heinz, Z.~Qiu, and C.~Shen, ``{Fluctuating flow angles and anisotropic flow
  measurements},'' \href{http://dx.doi.org/10.1103/PhysRevC.87.034913}{{\em
  Phys. Rev.} {\bfseries C87} no.~3, (2013) 034913},
\href{http://arxiv.org/abs/1302.3535}{{\ttfamily arXiv:1302.3535 [nucl-th]}}.

\bibitem{Aamodt:2011by}
{\bfseries ALICE} Collaboration, K.~Aamodt {\em et~al.}, ``{Harmonic
  decomposition of two-particle angular correlations in Pb-Pb collisions at
  $\sqrt{s_{NN}}=$ 2.76 TeV},''
  \href{http://dx.doi.org/10.1016/j.physletb.2012.01.060}{{\em Phys. Lett.}
  {\bfseries B708} (2012) 249--264},
\href{http://arxiv.org/abs/1109.2501}{{\ttfamily arXiv:1109.2501 [nucl-ex]}}.

\bibitem{Abelev:2012ola}
{\bfseries ALICE} Collaboration, B.~Abelev {\em et~al.}, ``{Long-range angular
  correlations on the near and away side in $p$-Pb collisions at
  $\sqrt{s_{NN}}=5.02$ TeV},''
  \href{http://dx.doi.org/10.1016/j.physletb.2013.01.012}{{\em Phys. Lett.}
  {\bfseries B719} (2013) 29--41},
\href{http://arxiv.org/abs/1212.2001}{{\ttfamily arXiv:1212.2001 [nucl-ex]}}.

\bibitem{ABELEV:2013wsa}
{\bfseries ALICE} Collaboration, B.~B. Abelev {\em et~al.}, ``{Long-range
  angular correlations of $\rm \pi$, K and p in p-Pb collisions at
  $\sqrt{s_{\rm NN}}$ = 5.02 TeV},''
  \href{http://dx.doi.org/10.1016/j.physletb.2013.08.024}{{\em Phys. Lett.}
  {\bfseries B726} (2013) 164--177},
\href{http://arxiv.org/abs/1307.3237}{{\ttfamily arXiv:1307.3237 [nucl-ex]}}.

\bibitem{Abelev:2013haa}
{\bfseries ALICE} Collaboration, B.~B. Abelev {\em et~al.}, ``{Multiplicity
  Dependence of Pion, Kaon, Proton and Lambda Production in p-Pb Collisions at
  $\sqrt{s_{NN}}$ = 5.02 TeV},''
  \href{http://dx.doi.org/10.1016/j.physletb.2013.11.020}{{\em Phys. Lett.}
  {\bfseries B728} (2014) 25--38},
\href{http://arxiv.org/abs/1307.6796}{{\ttfamily arXiv:1307.6796 [nucl-ex]}}.

\bibitem{Abelev:2014mda}
{\bfseries ALICE} Collaboration, B.~B. Abelev {\em et~al.}, ``{Multiparticle
  azimuthal correlations in p -Pb and Pb-Pb collisions at the CERN Large Hadron
  Collider},'' \href{http://dx.doi.org/10.1103/PhysRevC.90.054901}{{\em Phys.
  Rev.} {\bfseries C90} no.~5, (2014) 054901},
\href{http://arxiv.org/abs/1406.2474}{{\ttfamily arXiv:1406.2474 [nucl-ex]}}.

\bibitem{Aad:2012gla}
{\bfseries ATLAS} Collaboration, G.~Aad {\em et~al.}, ``{Observation of
  Associated Near-Side and Away-Side Long-Range Correlations in
  $\sqrt{s_{NN}}=$ 5.02 TeV Proton-Lead Collisions with the ATLAS Detector},''
  \href{http://dx.doi.org/10.1103/PhysRevLett.110.182302}{{\em Phys. Rev.
  Lett.} {\bfseries 110} no.~18, (2013) 182302},
\href{http://arxiv.org/abs/1212.5198}{{\ttfamily arXiv:1212.5198 [hep-ex]}}.

\bibitem{Aad:2013fja}
{\bfseries ATLAS} Collaboration, G.~Aad {\em et~al.}, ``{Measurement with the
  ATLAS detector of multi-particle azimuthal correlations in p+Pb collisions at
  $\sqrt{s_{NN}}=$ 5.02 TeV},''
  \href{http://dx.doi.org/10.1016/j.physletb.2013.06.057}{{\em Phys. Lett.}
  {\bfseries B725} (2013) 60--78},
\href{http://arxiv.org/abs/1303.2084}{{\ttfamily arXiv:1303.2084 [hep-ex]}}.

\bibitem{CMS:2013bza}
{\bfseries CMS} Collaboration, S.~Chatrchyan {\em et~al.}, ``{Studies of
  azimuthal dihadron correlations in ultra-central PbPb collisions at
  $\sqrt{s_{NN}} =$ 2.76 TeV},''
  \href{http://dx.doi.org/10.1007/JHEP02(2014)088}{{\em JHEP} {\bfseries 02}
  (2014) 088},
\href{http://arxiv.org/abs/1312.1845}{{\ttfamily arXiv:1312.1845 [nucl-ex]}}.

\bibitem{Chatrchyan:2013nka}
{\bfseries CMS} Collaboration, S.~Chatrchyan {\em et~al.}, ``{Multiplicity and
  transverse momentum dependence of two- and four-particle correlations in pPb
  and PbPb collisions},''
  \href{http://dx.doi.org/10.1016/j.physletb.2013.06.028}{{\em Phys. Lett.}
  {\bfseries B724} (2013) 213--240},
\href{http://arxiv.org/abs/1305.0609}{{\ttfamily arXiv:1305.0609 [nucl-ex]}}.

\bibitem{Chatrchyan:2013eya}
{\bfseries CMS} Collaboration, S.~Chatrchyan {\em et~al.}, ``{Study of the
  production of charged pions, kaons, and protons in pPb collisions at
  $\sqrt{s_{NN}} =\ $ 5.02 TeV},''
  \href{http://dx.doi.org/10.1140/epjc/s10052-014-2847-x}{{\em Eur. Phys. J.}
  {\bfseries C74} no.~6, (2014) 2847},
\href{http://arxiv.org/abs/1307.3442}{{\ttfamily arXiv:1307.3442 [hep-ex]}}.

\bibitem{Khachatryan:2014jra}
{\bfseries CMS} Collaboration, V.~Khachatryan {\em et~al.}, ``{Long-range
  two-particle correlations of strange hadrons with charged particles in pPb
  and PbPb collisions at LHC energies},''
  \href{http://dx.doi.org/10.1016/j.physletb.2015.01.034}{{\em Phys. Lett.}
  {\bfseries B742} (2015) 200--224},
\href{http://arxiv.org/abs/1409.3392}{{\ttfamily arXiv:1409.3392 [nucl-ex]}}.

\bibitem{Khachatryan:2015waa}
{\bfseries CMS} Collaboration, V.~Khachatryan {\em et~al.}, ``{Evidence for
  Collective Multiparticle Correlations in p-Pb Collisions},''
  \href{http://dx.doi.org/10.1103/PhysRevLett.115.012301}{{\em Phys. Rev.
  Lett.} {\bfseries 115} no.~1, (2015) 012301},
\href{http://arxiv.org/abs/1502.05382}{{\ttfamily arXiv:1502.05382 [nucl-ex]}}.

\bibitem{Aaij:2015qcq}
{\bfseries LHCb} Collaboration, R.~Aaij {\em et~al.}, ``{Measurements of
  long-range near-side angular correlations in $\sqrt{s_{\text{NN}}}=5$TeV
  proton-lead collisions in the forward region},''
  \href{http://dx.doi.org/10.1016/j.physletb.2016.09.064}{{\em Phys. Lett.}
  {\bfseries B762} (2016) 473--483},
\href{http://arxiv.org/abs/1512.00439}{{\ttfamily arXiv:1512.00439 [nucl-ex]}}.

\bibitem{Khachatryan:2015oea}
{\bfseries CMS} Collaboration, V.~Khachatryan {\em et~al.}, ``{Evidence for
  transverse momentum and pseudorapidity dependent event plane fluctuations in
  PbPb and pPb collisions},''
  \href{http://dx.doi.org/10.1103/PhysRevC.92.034911}{{\em Phys. Rev.}
  {\bfseries C92} no.~3, (2015) 034911},
\href{http://arxiv.org/abs/1503.01692}{{\ttfamily arXiv:1503.01692 [nucl-ex]}}.

\bibitem{Xu:2012ue}
L.~Xu, L.~Yi, D.~Kikola, J.~Konzer, F.~Wang, and W.~Xie, ``{Model-independent
  decomposition of flow and nonflow in relativistic heavy-ion collisions},''
  \href{http://dx.doi.org/10.1103/PhysRevC.86.024910}{{\em Phys. Rev.}
  {\bfseries C86} (2012) 024910},
\href{http://arxiv.org/abs/1204.2815}{{\ttfamily arXiv:1204.2815 [nucl-ex]}}.

\bibitem{Aamodt:2008zz}
{\bfseries ALICE} Collaboration, K.~Aamodt {\em et~al.}, ``{The ALICE
  experiment at the CERN LHC},''
\href{http://dx.doi.org/10.1088/1748-0221/3/08/S08002}{{\em JINST} {\bfseries
  3} (2008) S08002}.

\bibitem{Alme:2010ke}
J.~Alme {\em et~al.}, ``{The ALICE TPC, a large 3-dimensional tracking device
  with fast readout for ultra-high multiplicity events},''
  \href{http://dx.doi.org/10.1016/j.nima.2010.04.042}{{\em Nucl. Instrum.
  Meth.} {\bfseries A622} (2010) 316--367},
\href{http://arxiv.org/abs/1001.1950}{{\ttfamily arXiv:1001.1950
  [physics.ins-det]}}.

\bibitem{Aamodt:2010aa}
{\bfseries ALICE} Collaboration, K.~Aamodt {\em et~al.}, ``{Alignment of the
  ALICE Inner Tracking System with cosmic-ray tracks},''
  \href{http://dx.doi.org/10.1088/1748-0221/5/03/P03003}{{\em JINST} {\bfseries
  5} (2010) P03003},
\href{http://arxiv.org/abs/1001.0502}{{\ttfamily arXiv:1001.0502
  [physics.ins-det]}}.

\bibitem{Abelev:2014ffa}
{\bfseries ALICE} Collaboration, B.~B. Abelev {\em et~al.}, ``{Performance of
  the ALICE Experiment at the CERN LHC},''
  \href{http://dx.doi.org/10.1142/S0217751X14300440}{{\em Int. J. Mod. Phys.}
  {\bfseries A29} (2014) 1430044},
\href{http://arxiv.org/abs/1402.4476}{{\ttfamily arXiv:1402.4476 [nucl-ex]}}.

\bibitem{Abbas:2013taa}
{\bfseries ALICE} Collaboration, E.~Abbas {\em et~al.}, ``{Performance of the
  ALICE VZERO system},''
  \href{http://dx.doi.org/10.1088/1748-0221/8/10/P10016}{{\em JINST} {\bfseries
  8} (2013) P10016},
\href{http://arxiv.org/abs/1306.3130}{{\ttfamily arXiv:1306.3130 [nucl-ex]}}.

\bibitem{Abelev:2013qoq}
{\bfseries ALICE} Collaboration, B.~Abelev {\em et~al.}, ``{Centrality
  determination of Pb-Pb collisions at $\sqrt{s_{NN}}$ = 2.76 TeV with
  ALICE},'' \href{http://dx.doi.org/10.1103/PhysRevC.88.044909}{{\em Phys.
  Rev.} {\bfseries C88} no.~4, (2013) 044909},
\href{http://arxiv.org/abs/1301.4361}{{\ttfamily arXiv:1301.4361 [nucl-ex]}}.

\bibitem{ALICE:2012xs}
{\bfseries ALICE} Collaboration, B.~Abelev {\em et~al.}, ``{Pseudorapidity
  density of charged particles in $p$ + Pb collisions at $\sqrt{s_{NN}}=5.02$
  TeV},'' \href{http://dx.doi.org/10.1103/PhysRevLett.110.032301}{{\em Phys.
  Rev. Lett.} {\bfseries 110} no.~3, (2013) 032301},
\href{http://arxiv.org/abs/1210.3615}{{\ttfamily arXiv:1210.3615 [nucl-ex]}}.

\bibitem{Gyulassy:1994ew}
M.~Gyulassy and X.-N. Wang, ``{HIJING 1.0: A Monte Carlo program for parton and
  particle production in high-energy hadronic and nuclear collisions},''
  \href{http://dx.doi.org/10.1016/0010-4655(94)90057-4}{{\em Comput. Phys.
  Commun.} {\bfseries 83} (1994) 307},
\href{http://arxiv.org/abs/nucl-th/9502021}{{\ttfamily arXiv:nucl-th/9502021
  [nucl-th]}}.

\bibitem{Lin:2004en}
Z.-W. Lin, C.~M. Ko, B.-A. Li, B.~Zhang, and S.~Pal, ``{A Multi-phase transport
  model for relativistic heavy ion collisions},''
  \href{http://dx.doi.org/10.1103/PhysRevC.72.064901}{{\em Phys. Rev.}
  {\bfseries C72} (2005) 064901},
\href{http://arxiv.org/abs/nucl-th/0411110}{{\ttfamily arXiv:nucl-th/0411110
  [nucl-th]}}.

\bibitem{Brun:1994aa}
R.~Brun, F.~Bruyant, F.~Carminati, S.~Giani, M.~Maire, A.~McPherson,
  G.~Patrick, and L.~Urban, ``{GEANT Detector Description and Simulation
  Tool},''
{\em CERN-W} {\bfseries 5013} (1994) .

\bibitem{Roesler:2000he}
S.~Roesler, R.~Engel, and J.~Ranft,
  \href{http://dx.doi.org/10.1007/978-3-642-18211-2_166}{``{The Monte Carlo
  event generator DPMJET-III},''} in {\em {Advanced Monte Carlo for radiation
  physics, particle transport simulation and applications. Proceedings,
  Conference, MC2000, Lisbon, Portugal, October 23-26, 2000}}, pp.~1033--1038.
\newblock 2000.
\newblock \href{http://arxiv.org/abs/hep-ph/0012252}{{\ttfamily
  arXiv:hep-ph/0012252 [hep-ph]}}.
\newblock
\url{http://www-public.slac.stanford.edu/sciDoc/docMeta.aspx?slacPubNumber=SLAC-PUB-8740}.
\newblock

\bibitem{Zhao:2017yhj}
W.~Zhao, H.-j. Xu, and H.~Song, ``{Collective flow in 2.76 A TeV and 5.02 A TeV
  Pb+Pb collisions},''
\href{http://arxiv.org/abs/1703.10792}{{\ttfamily arXiv:1703.10792 [nucl-th]}}.

\bibitem{Qiu:2011hf}
Z.~Qiu, C.~Shen, and U.~Heinz, ``{Hydrodynamic elliptic and triangular flow in
  Pb-Pb collisions at $\sqrt{s}=2.76$ATeV},''
  \href{http://dx.doi.org/10.1016/j.physletb.2011.12.041}{{\em Phys. Lett.}
  {\bfseries B707} (2012) 151--155},
\href{http://arxiv.org/abs/1110.3033}{{\ttfamily arXiv:1110.3033 [nucl-th]}}.

\bibitem{Kozlov:2014fqa}
I.~Kozlov, M.~Luzum, G.~Denicol, S.~Jeon, and C.~Gale, ``{Transverse momentum
  structure of pair correlations as a signature of collective behavior in small
  collision systems},''
\href{http://arxiv.org/abs/1405.3976}{{\ttfamily arXiv:1405.3976 [nucl-th]}}.

\bibitem{Abelev:2013bla}
{\bfseries ALICE} Collaboration, B.~B. Abelev {\em et~al.}, ``{Multiplicity
  dependence of the average transverse momentum in pp, p-Pb, and Pb-Pb
  collisions at the LHC},''
  \href{http://dx.doi.org/10.1016/j.physletb.2013.10.054}{{\em Phys. Lett.}
  {\bfseries B727} (2013) 371--380},
\href{http://arxiv.org/abs/1307.1094}{{\ttfamily arXiv:1307.1094 [nucl-ex]}}.

\end{thebibliography}\endgroup
